\documentclass[pre,twocolumn,showkeys,english,aps,a4paper,floatfix]{revtex4}
\usepackage[T1]{fontenc}
\usepackage{color}
\usepackage{graphicx}
\usepackage{amsmath}
\usepackage{amssymb}
\usepackage[latin1]{inputenc}
\usepackage{graphicx}
\usepackage{bm}
\usepackage{epsfig}
\usepackage{morefloats}
\usepackage{float}
\begin{document}
\maxdeadcycles=500

\providecommand{\keywords}[1]
{
  \small
  \textbf{\textit{Keywords---}} #1
}

\title{Highly confined mixtures of parallel hard squares: A Density Functional Theory study}

\author{Yuri Mart\'{\i}nez-Rat\'on}
\email{yuri@math.uc3m.es}

\address{Grupo Interdisciplinar de Sistemas Complejos (GISC), Departamento de Matem\'aticas, 
Escuela Polit\'ecnica Superior, Universidad Carlos III de Madrid, Avenida de la Universidad 30, 
E-28911, Legan\'es, Madrid, Spain}

\author{Enrique Velasco}
\email{enrique.velasco@uam.es}

\address{Departamento de F\'{\i}sica Te\'orica de la Materia Condensada, Instituto de F\'{\i}sica 
de la Materia Condensada (IFIMAC) and Instituto de Ciencia de Materiales Nicol\'as Cabrera, 
Universidad Aut\'onoma de Madrid, E-28049, Madrid, Spain}

\begin{abstract}
Using the Fundamental-Measure Density Functional Theory, we have studied theoretically the phase behavior of 
extremely confined mixtures of parallel hard squares in slit geometry. The pore width is chosen such that
configurations consisting of two consecutive big squares, or three small squares, in the transverse 
direction, perpendicular to the walls, are forbidden. We analyzed two different mixtures with edge-lengths 
of species selected so as to allow or forbid one big plus one small square to fit into the channel. For the 
first mixture we obtained first-order transitions between symmetric and asymmetric packings of particles: 
small and big squares are preferentially adsorbed at different walls. Asymmetric configurations are shown to lead to more efficient packing
at finite pressures. We argue that the stability region of the asymmetric phase in the 
pressure-composition plane is bounded so that the symmetric phase is stable at low and very high pressure.
For the second mixture, we observe strong demixing between phases which are rich in different species.
Demixing occurs in the transverse direction, i.e. the dividing interface is perpendicular to the walls,
and phases exhibit symmetric density profiles. The possible experimental 
realization of this behaviour (which in practical terms is precluded by jamming)
in strictly two-dimensional systems is discussed.
Finally the phase behavior of a mixture with periodic boundary 
conditions is analyzed and the differences and similarities between the latter and the confined system are
discussed. We claim that, although exact calculations discard the existence of true phase transitions 
in $1+\epsilon$-dimensional systems, Density Functional Theory is still successful to describe 
packing properties of large clusters of particles.  
\end{abstract}

\date{\today}


\maketitle

\section{Introduction}
\label{intro}
Fluids of two-dimensional hard anisotropic particles are paradigmatic examples of systems 
exhibiting entropy-driven phase transitions to orientationally and positionally ordered phases. 
The elucidation of the phase behavior of two-dimensional fluids composed of hard particles 
is not an academic study, since hard particles enjoy many experimental realizations.
To cite a representative recent experiment, extreme confinement of three-dimensional lithographically 
synthesized prisms with different polygonal cross-sections in quasi-2D geometries has been accomplished 
\cite{Zhao,Wang,Rossi,Qi}. The phase behavior of these effectively two-dimensional Brownian particles 
were reported, and their tendency to produce chiral phases \cite{Zhao,Rossi} or racemic mixtures of 
monomers and dimers was emphasized \cite{Wang}. Also, the phase behavior of colloidal monolayers of 
particles with exotic shapes has recently been reported \cite{Qi}. Research on three-dimensional
colloidal particles of different shapes, especially in connection with packing and partial or
complete crystallization, has also been very active (see Ref. \cite{Manoharan} for a recent review).
		
Several theoretical works have concentrated on the elucidation of the phase behavior of hard polygonal particles 
\cite{Frenkel,M-R1,Donev,Avendano,Gantapara,M-R2,Anderson,Shen,Thapar}. The results show that it strongly 
depends on the symmetries of particle shapes. Apart from the usual uniaxial nematic phase present in fluids of 
elongated rods, other more `exotic' orientational fluid phases, such as triatic, tetratic and hexactic phases,
also exist. For example, hard rectangles may order into uniaxial nematic phases, but also in
tetratic arrangements at low particle aspect ratios \cite{Donev,M-R1}. Different plastic or orientationally 
ordered crystals have been classified as a function of particle shape \cite{Anderson,Shen}. 
Especially interesting is the case of hard squares because of their 
plane-filling properties and the mathematical simplicity of their interaction potential.
Classical work on the numerical calculation of virial coefficients \cite{Hoover1,Hoover2}
demonstrate the importance
of hard squares as a simple model to elucidate important problems in statistical mechanics.
The lattice-gas version of the model has attracted some attention \cite{Lafuente,Ramola,Singh}
The parallel hard square model has also been investigated \cite{Hoover3,Belli,Pinto}.
Simulations have shown that freely-oriented hard squares present 
nematic tetratic and crystal square phases \cite{Frenkel}. Rounded hard squares
have been investigated and their phase behaviours seen to depend 
on the degree of roundness \cite{Avendano}. An experimental realization of
this system has been reported, together with evidence for a rich phase
diagram \cite{Zhao1}. Also, demixing transitions in mixtures of hard squares have been
explored by simulation \cite{Buhot}.

The effect of confinement on two-dimensional fluids of rod-like particles in cavities of square, 
rectangular or circular geometries has been extensively studied 
\cite{Heras2,Heras3,Geigenfeind,Garlea,Lewis,Manyuhina,Heras4,Gonzalez-Pinto}. 
When the confining geometry is incompatible with the symmetry of the bulk phase the system usually
responds to the geometric frustration by creating point defects or domain walls in the orientational field.
Hard particles exhibit preferred orientations at the boundary of the confining walls, which are controlled
solely by entropy. These `anchoring' effects are strong enough that creation of defects is unavoidable.
The number and symmetry of the defects strongly depends on the geometry of the confining cavity and on the 
symmetries of the bulk phases.

When confinement of 2D hard particles (an also of 3D hard spheres inside a cylindrical pore) 
between two hard lines is so extreme that the system is close to the 1D limit 
the partition function can be calculated for nearest-neighbor or next-nearest-neighbor interactions 
using the Transfer Matrix Method (TMM).
This method becomes a useful (and potentially exact) theoretical tool to extract information about the structure of the 
confined fluid. In essence, the technique calculates, apart from the partition function, the probability density 
and pair correlations between particles. The method was successfully applied to the study of hard disks, 
squares, rhombuses and rectangles under strong confinement 
\cite{Kofke,Gurin5,Gurin4,Godfrey,Gurin2,Gurin1,Gurin3,Hu}. The results can be summarized as follows: 
(i) Phase transitions between different spatial structures are ruled out, a confirmation of the general result 
that fluids composed of particles interacting via hard-core potentials do not exhibit phase transitions 
in $1+\epsilon$ dimensions. (ii) From the behavior of probability densities and pair correlation functions
smooth crossovers between different spatial structures can be shown to exist. 
For example, the system may change from 
a one-layer structure that behaves approximately as a 1D Tonk's gas to a structure consisting of two 
highly-correlated layers adsorbed at each wall. Correlation may be different depending on the specific 
particle geometry (circular vs. square). (iii) Although phase transitions can be ruled out, the equation 
of state (EOS) may exhibit, in a range of packing fraction associated with the structural crossover, 
a plateau, and consequently the specific heat exhibits a sharp peak in this range of packing fraction.

The implementation of the TMM to such systems serves as an ideal testbed to study the performance 
of available Density Functionals (DF) developed for 2D fluids of: hard disks \cite{Roth}, 
parallel hard squares (PHS) \cite{cuesta1}, rectangles within the restricted orientation approximation 
\cite{cuesta2}, or freely-rotating disco-rectangles \cite{Wittmann}. All of these DF are based on the 
Fundamental-Measure Theory (FMT), initially developed for hard spheres and further extended to 
anisotropic particles. For reviews of this theoretical tool see Refs. \cite{Roth2,Tarazona}. 
Recent work on highly confined PHS and rectangles (in the orientation-restricted or Zwanzig approximation) 
in slit geometry using both theories, TMM and FMT, demonstrated the high performance of FMT to predict 
changes in the structural properties of the fluid induced by confinement, and also to describe the 
anomalous behavior of the EOS at the crossover between different structures \cite{Gurin3,pinto1}.

In the present article we go beyond the one-component fluid studied previously, and focus on 
the effect of extreme confinement on the structural and thermodynamical properties of binary mixtures of PHS,
using a FMT-based formalism. Mixtures of small (edge-length equal to $\sigma_1$)
and big (edge-length equal to $\sigma_2$) squares are confined into a channel of width $H$.
The value of $H$ is selected in such a way that at most two layers of small squares can fit into the channel, 
whereas only one layer (but not two) of big squares can fit. We analyze two different mixtures characterized by 
the ratios $\sigma_2/\sigma_1=1.5$ and 2. We found micro- and macrosegregation   
first-order transitions for the first and second mixtures, respectively. In the former case, 
different species are preferentially adsorbed at different walls, while in the latter species phase-separate, 
with a dividing surface perpendicular to the walls. We explain, using entropic arguments, why these mixtures 
segregate. We claim that a TMM applied to these mixtures could confirm the appearance of large clusters of 
micro/macrosegregated particles as the packing fraction is increased, despite the fact that an exact theory 
should discard the existence of a true phase transition between different structures.  

The paper is organized as follows: In Sec. \ref{model} the model is introduced 
and details are provided on the theory. Also, the numerical procedure 
used to find the phase behavior of the system is discussed.
Technical details to prove the nonexistence of fluid-fluid demixing at bulk, along with 
the method to find the spinodal instability of uniform phases with respect to 1D spatial density 
modulations, are relegated to Sec. \ref{uniform} and \ref{app_sf}, respectively. 
In Sec. \ref{results} the results are presented. This section is in turn divided into two parts,
where results obtained for mixtures with $\sigma_2/\sigma_1=1.5$ 
[Sec. \ref{tres_medio_uno}] and $\sigma_2/\sigma_1=2$ [Sec. \ref{dos_uno}] are given.
The end of Sec. \ref{dos_uno} is devoted to describing the 
phase behavior of the $\sigma_2/\sigma_1=2$ mixture that results from
imposing periodic boundary conditions, 
instead of a confining external potential. 
Finally some conclusions are drawn in Sec. \ref{conclusions}. 

\begin{figure}[H]
	\epsfig{file=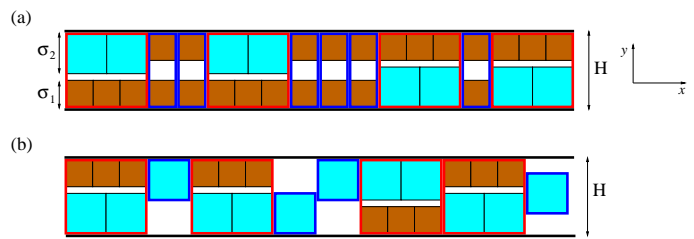,width=3.5in}
	\caption{Schematic of close-packing configurations of binary mixtures of PHS with  
	$\sigma_2/\sigma_1=1.5$ and molar fractions (a) ${\sf x}>3/5$, and (b) ${\sf x}<3/5$. The small and big 
	clusters are indicated with blue and red solid lines, respectively.}
	\label{fig0_new}
\end{figure}

\section{Model and Theory}
\label{model}
Our model consists of a binary mixture of PHS confined into a channel (or a slit pore)  
formed by two parallel hard lines (or walls) with a relative distance between them 
of $H$ (the pore width). See Fig. \ref{fig0_new} 
for a sketch of the system. Small and large particles have edge-lengths equal to 
$\sigma_1$ and $\sigma_2$, respectively. Coordinates parallel and 
perpendicular to the walls are chosen as $x$ and $y$, respectively, and the walls are 
located at $y=0$ and $y=H$. Our system is described in terms of the 
density profile of species $i$, $\rho_i(y)$, which is assumed to depend only on the $y$-coordinate. 
The mean density, averaged in the channel, of the $i$th species is defined as 
\begin{eqnarray}
\rho_i\equiv\frac{1}{H}\int_0^H dy \rho_i(y),\quad \rho=\rho_1+\rho_2, 
\end{eqnarray}
with $\rho$ the total mean density. The mixture composition is described in terms of the mean 
molar fraction of small species:
\begin{eqnarray}
	{\sf x}\equiv {\sf x}_1=\frac{\rho_1}{\rho},\quad {\sf x}_2=\frac{\rho_2}{\rho}=1-{\sf x},
	\quad \sum_i {\sf x}_i=1.
\end{eqnarray}
The mean packing fraction of the mixture is, as usual, calculated as   
\begin{eqnarray}
	\eta=\sum_{i=1}^2 \eta_i=\sum_{i=1}^2\rho_i\sigma_i^2.
\end{eqnarray}
The theoretical model used is a version of DFT, the so-called FMT, which was formulated 
for PHS in the '90 \cite{cuesta1} and has been extensively tested before in several studies 
\cite{Gurin3,pinto1}. The main assumption of the 
theory, adapted to the present system, is that the excess (or interaction) part of the 
free-energy density of the PHS fluid only depends on four weighted densities,
\begin{eqnarray}
	&&n_0(y)=\frac{1}{2}\sum_i \left[\rho_i(y_i^{-})+\rho_i(y_i^+)\right],\\
	&&n_2(y)=\sum_i\sigma_i\int_{y_i^-}^{y_i^+}dy'\rho_i(y'),\\
	&&n_{1x}(y)=\frac{1}{2}\sum_i\sigma_i\left[\rho_i(y_i^{-})+\rho_i(y_i^+)\right],\\
	&&n_{1y}(y)=\sum_i\int_{y_i^-}^{y_i^+}dy'\rho_i(y'),
\end{eqnarray}
where $y_i^{\pm}=y\pm\sigma_i/2$. The explicit expression for the excess free-energy 
density, in reduced thermal units, is \cite{cuesta1}
\begin{eqnarray}
	\Phi_{\rm exc}(y)=-n_0(y)\log[1-n_2(y)]+\frac{n_{1x}(y)n_{1y}(y)}{1-n_2(y)},
\end{eqnarray}
while the ideal part, neglecting the thermal areas, is 
\begin{eqnarray}
	\Phi_{\rm id}(y)=\sum_i \rho_i(y)\left[\log \rho_i(y)-1\right],\\
\end{eqnarray}
The grand-potential per unit length can then be calculated as 
\begin{eqnarray}
	\frac{\Omega[\{\rho_i\}]}{L}=\frac{{\cal F}[\{\rho_i\}]}{L}-\sum_i \int_0^H dy  
	\left(\mu_i-v^{(i)}_{\rm ext}(y)\right)\rho_i(y), 
\end{eqnarray}
with ${\cal F}[\{\rho_i\}]$ the Helmholtz free-energy DF,
\begin{eqnarray}
	&&\frac{\beta {\cal F}[\{\rho_i\}]}{L}=\frac{\beta {\cal F}_{\rm id}[\{\rho_i\}]}{L}+
	\frac{\beta {\cal F}_{\rm exc}[\{\rho_i\}]}{L}\nonumber\\
	&&=\int_0^H dy \Phi_{\rm id}(y)+
	\int_0^H dy\Phi_{\rm exc}(y),
\end{eqnarray}
with $\beta=(k_B T)^{-1}$ the inverse of temperature, 
$\mu_i$ the chemical potential of species $i$ and $L$ the length of the system. 
The external potential acting on particle $i$ 
is defined as
\begin{eqnarray}
	\beta v_{\rm ext}^{(i)}(y)=\left\{
		\begin{matrix}
			0, & \displaystyle{\frac{\sigma_i}{2}\leq y\leq H-\frac{\sigma_i}{2}},\\
                        \infty, & \text{otherwise.}
		\end{matrix}
		\right.
\end{eqnarray}
By minimizing the grand potential with respect to $\rho_i(y)$, i.e. 
$\displaystyle{\frac{\delta \beta \Omega[\{\rho_i\}]}{\delta \rho_i(y)}=0}$, 
we obtain 
\begin{eqnarray}
	\rho_i(y)=\left\{
		\begin{matrix}
			e^{-\Psi_i(y)+\beta\mu_i}, & \displaystyle{\frac{\sigma_i}{2}\leq y\leq H-\frac{\sigma_i}{2}},\\
			0, & \text{otherwise,}
		\end{matrix}
		\right.
		\label{solve1}
\end{eqnarray}
where we have used the short-hand notation  
\begin{eqnarray}
	&&\Psi_i(y)\equiv \frac{\delta \beta {\cal F}_{\rm exc}[\{\rho_i\}]/L}{\delta \rho_i(y)}
\end{eqnarray}
The longitudinal pressure inside the channel can be calculated as
\begin{eqnarray}
	\beta p&=&\frac{1}{H}\left\{\sum_i\left[\int_0^H dy \rho_i(y)\left(1+\Psi_i(y)\right)\right]
	-\frac{\beta {\cal F}_{\rm exc}}{L}\right\}\nonumber\\
	&=&\frac{1}{H}
	\int_0^Hdy\left[\frac{n_0(y)}{1-n_2(y)}+\frac{n_{1x}(y)n_{1y}(y)}{(1-n_2(y))^2}\right].
\end{eqnarray}
By fixing the values of both mean packing fractions $\eta_i$ inside the channel,
the constrained minimization of the free-energy,
$\beta {\cal F}[\{\rho_i\}]$, with respect to $\rho_i(y)$ leads to
\begin{eqnarray}
        \rho_i(y)=\frac{\eta_ie^{-\Psi_i(y)}}
        {\sigma_i^2H^{-1}\int_0^H dy' e^{-\Psi_i(y')}},
        \label{solve2}
\end{eqnarray}
for $\sigma_i/2\leq y\leq H-\sigma_i/2$, and zero otherwise. Obviously the two routes: (i) to fix 
the chemical potentials $\mu_i$, and (ii) to fix the packing fractions $\eta_i$, are equivalent. 
Using the second route to calculate the equilibrium density profiles, 
the chemical potentials can be calculated as 
\begin{eqnarray}
	\beta\mu_i=\log\left[\frac{\eta_i}{\sigma_i^2H^{-1}\int_0^Hdy e^{-\Psi_i(y)}}\right].
	\label{chepocon}
\end{eqnarray}
To study the thermodynamics of the confined fluid mixture, which is necessary to calculate
possible phase transitions, it is more convenient
to use the Gibbs free-energy per-particle in reduced thermal units, defined as 
\begin{eqnarray}
	g\equiv \frac{\beta}{\rho}\left(\frac{{\cal F}}{LH}+p_0\right).
	\label{elgibbs}
\end{eqnarray}
Here the pressure of the confined mixture is fixed,
\begin{eqnarray}
	p\left({\sf x},\rho\right)=p_0,
	\label{solve3}
\end{eqnarray}
and $\rho$ can be numerically calculated as a function of the mixture composition 
${\sf x}$ once the equilibrium 
density profiles, $\{\rho_i^{(\rm eq)}(y)\}$ are obtained from Eq. (\ref{solve2}). 
The function $g({\sf x})$ can then be obtained. In case of first-order phase transitions
a double-tangent construction on $g({\sf x})$ allows us to calculate the coexisting 
values of molar and packing fractions. For convenience we will use a dimensionless
pressure $p_0^*\equiv \beta p_0\sigma_1^2$.

Sec. \ref{uniform} of the Appendix presents a proof that the uniform mixture of PHS 
is always stable at bulk, i.e.  no demixing is possible.  In Sec. \ref{app_sf} of the same
Appendix the spinodal instability of uniform phases with respect to one-dimensional periodic 
inhomogeneities is discussed by means of a bifurcation analysis.

\section{Results}
\label{results}
This section is devoted to presenting the results obtained from the numerical solutions 
of Eqs. (\ref{solve2}) and (\ref{solve3}), which provide the equilibrium density profiles 
$\rho_i(y)$ for fixed pressure $p_0^*$, and for a given 
mixture composition ${\sf x}$. Varying ${\sf x}$ inside a given set of values
$\{{\sf x}_i=i/N_{\sf x},\ i=0,\dots,N_{\sf x}, \ N_{\sf x}\sim 100\}$ allows to obtain
a sufficiently accurate Gibbs free-energy per particle, $g({\sf x})$ [from Eq. (\ref{elgibbs})]
to search for possible phase transitions and
calculate the phase diagrams. This section is divided into two parts. Sec. \ref{tres_medio_uno} 
is concerned with a confined binary mixture of PHS with $\sigma_2/\sigma_1=1.5$ and several values 
of pore width $H/\sigma_1$. Values of $H$ were chosen to ensure that only two small squares 
(but not three) or one big plus one small square can fit 
inside the channel along its transverse direction, whereas only one big square (but not two) 
is allowed to fit (i.e. $2.5=1+\sigma_2/\sigma_1<H/\sigma_1<2\sigma_2/\sigma_1=3$). In Sec. 
\ref{dos_uno} a mixture with $\sigma_2/\sigma_1=2$ is studied. This time configurations
where one big plus one small or, again two big squares, are both forbidden, which is
expressed by the condition $2<H/\sigma_1<1+\sigma_2/\sigma_1=3$. 

\begin{figure}
        \epsfig{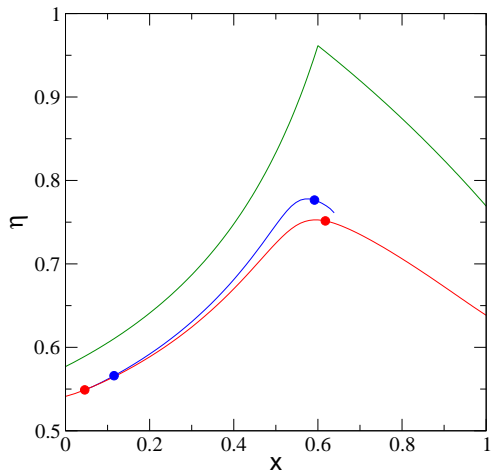}
	\caption{Packing fraction $\eta$ vs. molar fraction ${\sf x}$ for SYM (red) and ASYM (blue) 
	configurations of a confined binary mixture of PHS with $\sigma_2/\sigma_1=1.5$, 
        $H/\sigma_1=2.6$ and $p_0^*=4$. The green curve
corresponds to the packing fraction for the close-packed configuration $\eta_{\rm cp}$ 
	(see the text).
	Note that the maxima of the curves $\eta({\sf x})$ in both 
	SYM and ASYM configurations are located close to ${\sf x}=0.6$, the maximum close-packing 
	value. Red and blue circles indicate the SYM and ASYM coexisting states, respectively. 
Thus the red curve between red circles corresponds to metastable states. These states also occur 
in the blue curve at left and right of the blue circles.}
	\label{fig3}
\end{figure}

\subsection{The $\sigma_2/\sigma_1=3/2$ mixture}
\label{tres_medio_uno}

First we analyze the close-packing properties of the mixture. 
For composition ${\sf x}\geq 3/5$ the close-packing configuration can be reached 
by adding up along the channel two kinds of clusters in close contact. Big clusters,
${\cal N}_{\rm b}$ in number, consist of groups of five particles: two big squares 
joined in the direction along the channel and  in contact with one wall, and three small squares, 
also joined along the channel, located on top of (or below) the big squares and occupying 
the same length, parallel to the walls, as the big squares.
The other, smaller clusters, ${\cal N}_{\rm s}$ in number, are made of small squares grouped together
in dimers and consist of two squares, perfectly aligned along the transverse direction, 
each one in contact with opposite walls. See Fig. \ref{fig0_new} for a sketch of a
possible close-packing configuration. The number of clusters should fulfill 
the relation $3{\cal N}_{\rm b}+2{\cal N}_{\rm s}={\sf x} {\cal N}$ and $2{\cal N}_{\rm b}=(1-{\sf x}){\cal N}$ 
(and thus ${\cal N}_{\rm b}=(1-{\sf x}){\cal N}/2$ 
and ${\cal N}_{\rm s}=(5{\sf x}-3){\cal N}/4$) with ${\cal N}$ the total number of particles. The 
packing fraction at close packing can be calculated as the ratio between the total area 
occupied by all clusters divided by the total area, i.e.
\begin{eqnarray}
	\eta_{\rm cp}=\frac{{\cal N}_{\rm b}(2\sigma_2^2+3\sigma_1^2)+2{\cal N}_{\rm s}\sigma_1^2}
	{\left[3{\cal N}_{\rm b}\sigma_1+{\cal N}_{\rm s}\sigma_1\right]H}
	=\frac{9-5{\sf x}}{3-{\sf x}}\times\frac{\sigma_1}{H}, 
	\label{cp1}
\end{eqnarray}
for $\displaystyle{{\sf x}\geq 3/5}$.
For the case ${\sf x}\leq 3/5$ the close packing configuration can be reached by adding in close contact 
along the channel the same big clusters as defined previously, with a total amount of ${\cal N}_{\rm b}$, 
and small clusters, with a total number of ${\cal N}_{\rm s}$, this time formed by a single big square 
in any position along the transverse direction. The
numbers $\{{\cal N}_{\rm b},{\cal N}_{\rm s}\}$ fulfills 
$3{\cal N}_{\rm s}={\sf x} {\cal N}$ and $2{\cal N}_{\rm s}+{\cal N}_{\rm b}=(1-{\sf x}){\cal N}$ (and 
consequently ${\cal N}_{\rm s}={\sf x} {\cal N}/3$ and ${\cal N}_{\rm b}=(3-5{\sf x}){\cal N}/3$ .  
Then the packing fraction at close packing can be calculated for $\displaystyle{{\sf x}\leq 3/5}$ as
\begin{eqnarray}
	\eta_{\rm cp}=\frac{{\cal N}_{\rm b}(2\sigma_2^2+3\sigma_1^2)+{\cal N}_{\rm s}\sigma_2^2}
	{\left(3{\cal N}_{\rm b}\sigma_1+{\cal N}_{\rm s}\sigma_2\right)H}=\frac{9-5{\sf x}}{6(1-{\sf x})}
	\times \frac{\sigma_1}{H}. 
	\label{cp2}
\end{eqnarray}

\begin{figure}
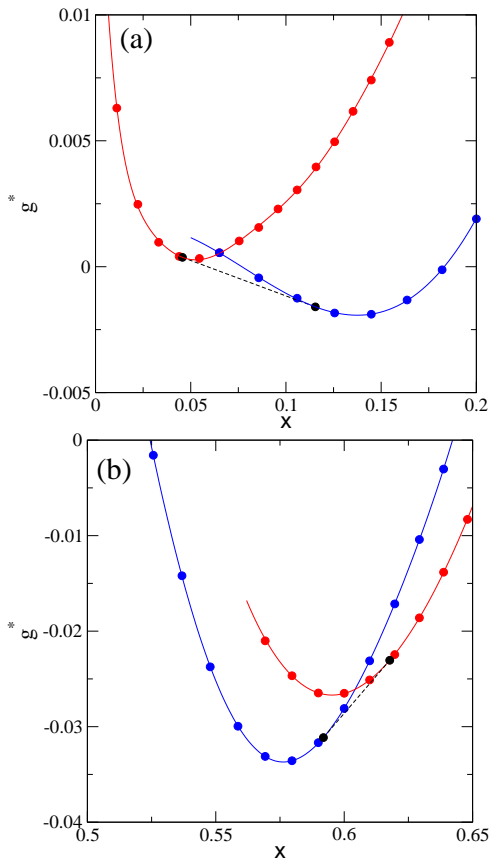

	\epsfig{file=Fig3a.eps,width=2.5in}
	\epsfig{file=Fig3b.eps,width=2.5in}
	       \caption{
	Gibbs free-energy per particle in reduced thermal units minus a straight line 
	vs. mean molar fraction ${\sf x}$. 
	(a) $g^*\equiv g-17.856+13.184 {\sf x}$ and (b) 
	$g^*\equiv g-15.959+8.873 {\sf x}$. Two different intervals of ${\sf x}$ are shown,
	located where transitions from SYM (red curve) to ASYM (blue 
	curve) [shown in (a)] and from ASYM to SYM [shown in (b)] phases take place. 
	Results correspond to a confined binary mixture of PHS with $\sigma_2/\sigma_1=1.5$, 
	$H/\sigma_1=2.6$ and $p_0^*=4$. Solid curves are least-square  
	polynomial fits to the red and blue symbols, which represent the calculated points. 
        The coexisting points are indicated with black symbols joined by dashed lines.}
\label{fig2}
\end{figure}

\begin{figure*}
        \epsfig{file=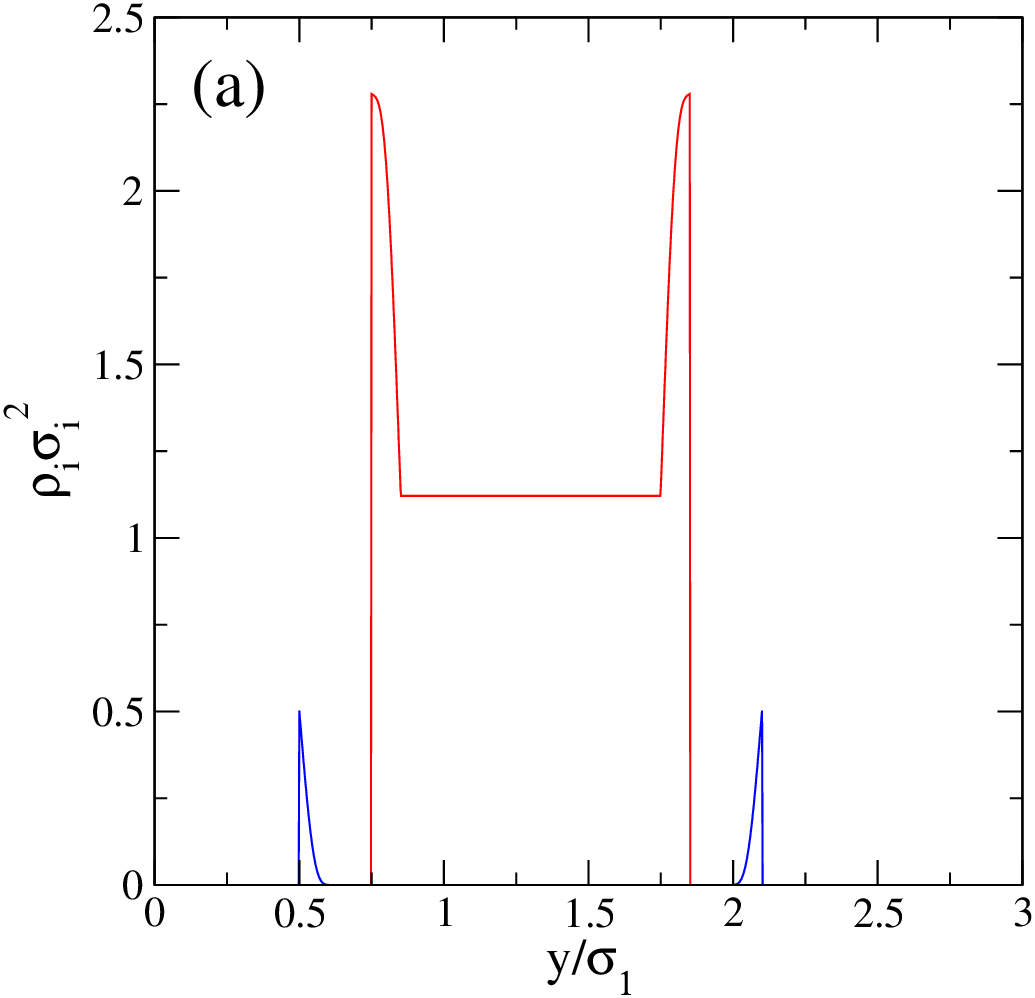,width=1.6in}
        \epsfig{file=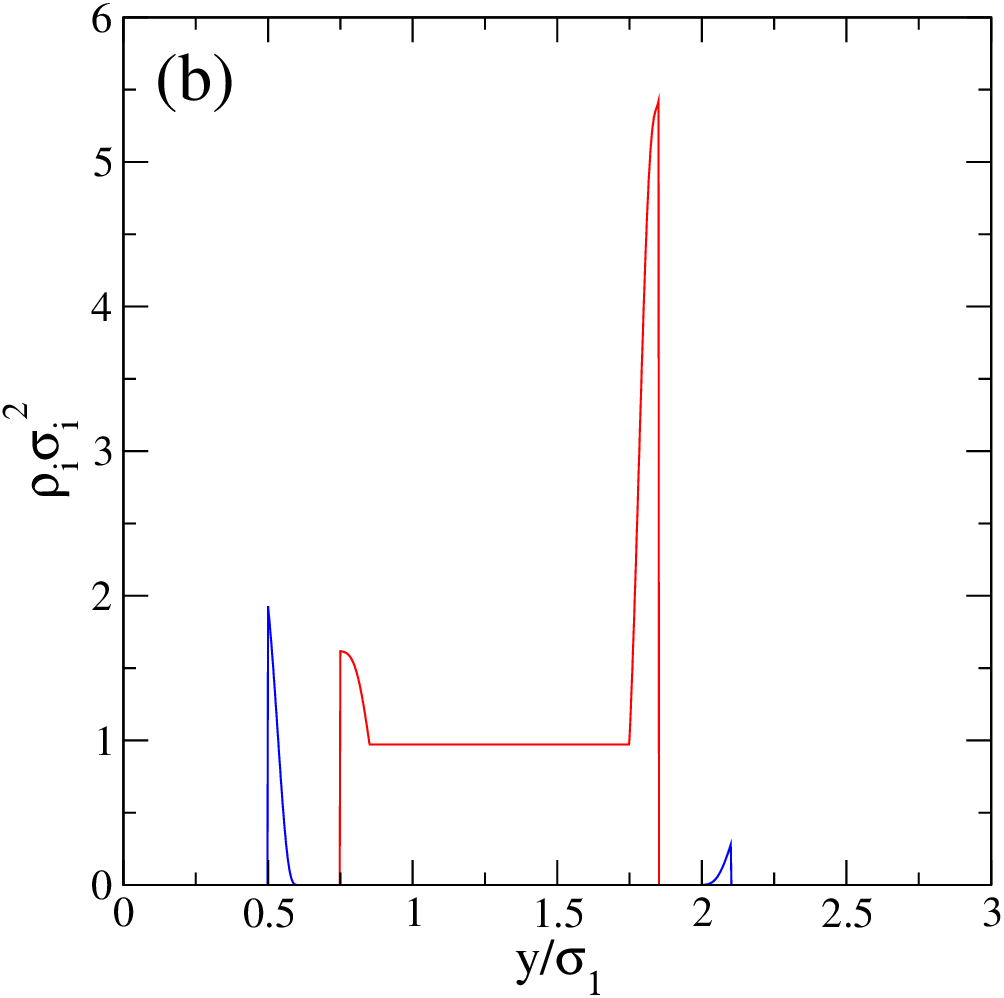,width=1.6in}
        \epsfig{file=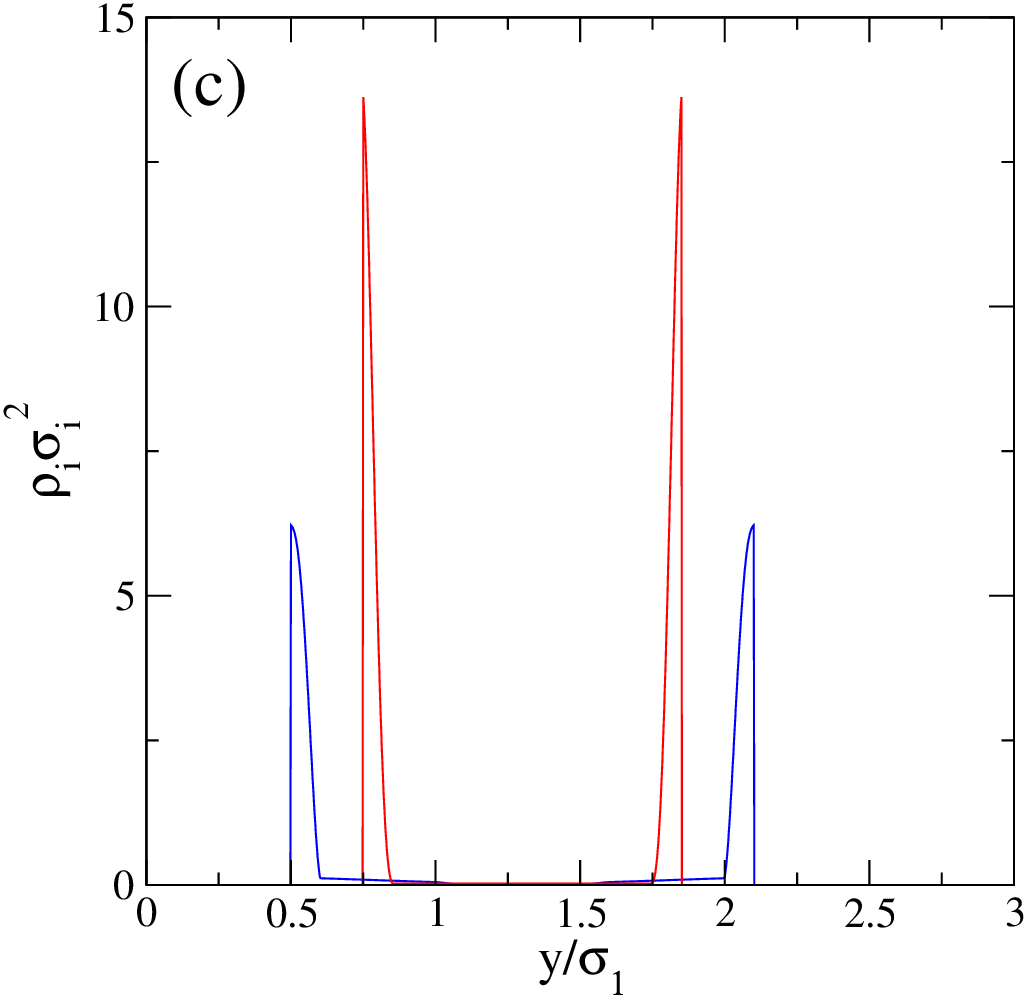,width=1.6in}
        \epsfig{file=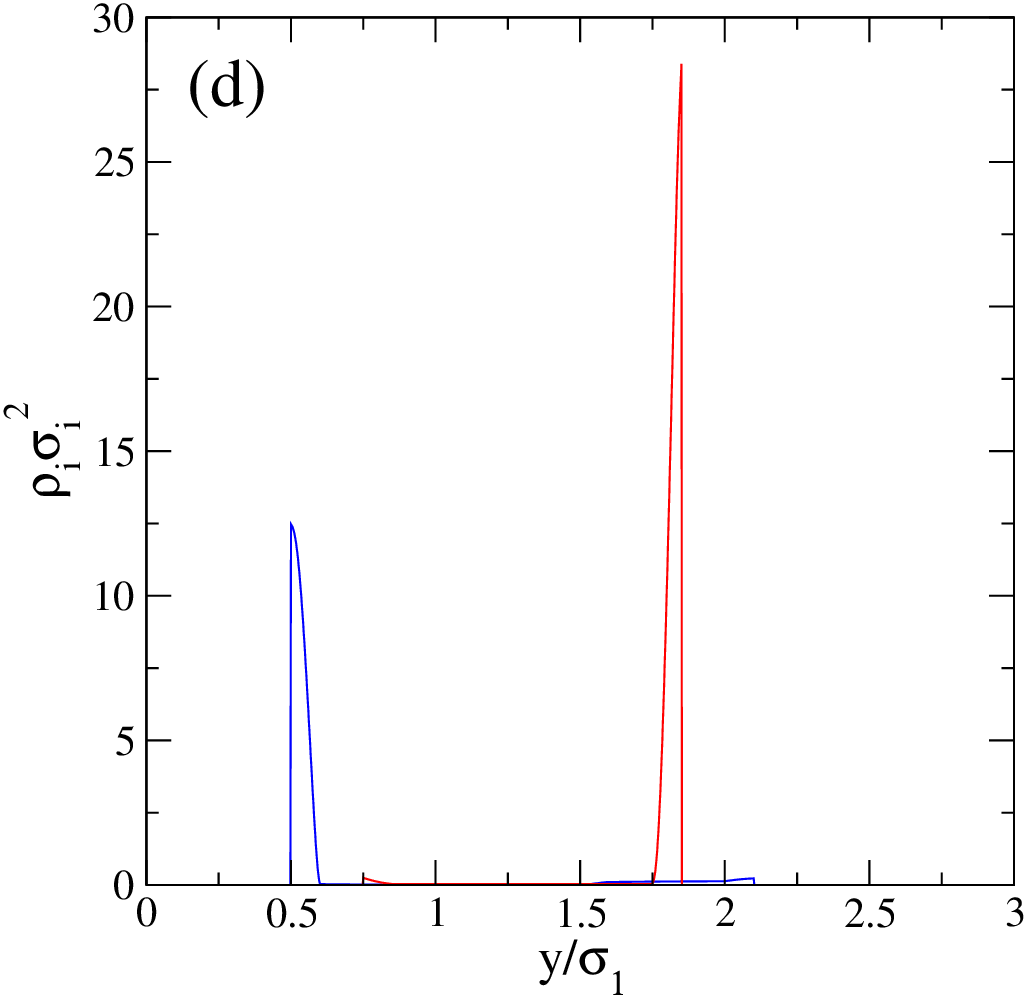,width=1.6in}
	\caption{Coexistence density profiles corresponding to the two first-order phase transitions shown
in Fig. \ref{fig2} for $\sigma_2/\sigma_1=1.5$, $H/\sigma_1=2.6$ and $p_0^*=4$. (a) and (b) correspond
to the low molar fraction region, whereas (c) and (d) refer to the high molar fraction region.
(a) and (c) are SYM phases, while (b) and (d) are ASYM phases. Blue and red curves 
	correspond to scaled density profiles $\rho_1(y)\sigma_1^2$ and $\rho_2(y)\sigma_2^2$ of 
	small and big species, respectively.} 
	\label{fig4}
\end{figure*}
The function $\eta_{\rm cp}({\sf x})$, given by Eqs. (\ref{cp1}) and (\ref{cp2}) for
the case $H/\sigma_1=2.6$, is plotted in green in Fig. \ref{fig3}. The maximum packing fraction 
obviously corresponds to ${\sf x}=3/5$ with 
$\displaystyle{\eta_{\rm cp}^{(\rm max)}=\eta_{\rm cp}(3/5)=5\sigma_1/2H}$.
The packing fractions for the one-component fluids composed of big and small 
particles are respectively $\displaystyle{\eta_{\rm cp}(0)= 3\sigma_1/2H}$ and 
$\displaystyle{\eta_{\rm cp}(1)=2\sigma_1/H}$. In the same figure we plot the results 
from DFT calculations for the same pore-width $H/\sigma_1=2.6$ 
at a fixed pressure $p_0^*=4$. Two different solutions are obtained, corresponding 
to two different local minima of the Gibbs free-energy per particle. The red line represents the so-called 
symmetric (SYM) solution, which has density profiles symmetric with respect to a line parallel to the
$x$ axis that passes through the middle of the channel, i.e. $\rho_i(y)=\rho_i(H-y)$.
The blue line, in contrast, represent an asymmetric (ASYM) solution, with $\rho_i(y)\neq \rho_i(H-y)$.
Note that the ASYM solution only exists in a particular interval of molar fractions,
whereas the SYM profile exists for all ${\sf x}$. The former gives a higher value of mean packing 
fraction $\eta$ (since the blue curve is above the red one). Both curves 
have their maxima located close to ${\sf x}\approx 3/5$, where the maximum value at close-packing 
is reached. The confined mixture exhibits two first-order phase transitions that take place as the molar 
fraction is increased from 0 to 1. 
Both the SYM-ASYM and ASYM-SYM transitions are correspondingly labeled in Fig. \ref{fig3}, with 
the coexisting values shown by two pairs of red and blue circles. 

The ASYM-phase 
is stable in an interval of ${\sf x}$ between the blue circles of Fig. \ref{fig3}. This can be concluded from   
Fig. \ref{fig2}, where we plot the Gibbs free-energy per particle 
for two different ranges of ${\sf x}$ [(a) and (b)] located close to both
phase transitions. In both cases straight lines have been subtracted to improve visualization.
The circles correspond to values of ${\sf x}$ where DFT calculations were performed, and the 
pressure was fixed to $p_0^*=4$. Red and blue curves are polynomial fits of the SYM and ASYM 
solutions respectively, which were used to calculate coexistence through a double-tangent construction. 
We checked that the energy $g$ of the ASYM-phase is always below that of SYM-phase in the interval
${\sf x}\in[0.2,0.5]$. The four coexisting density profiles are shown in Fig. \ref{fig4}. Their symmetric
 or asymmetric character are quite apparent. In the  ASYM-phase big and small squares are preferentially 
absorbed at different walls, a type of microsegregation transition. In contrast, in the SYM phase 
both species are equally adsorbed at both walls. 

\begin{figure}
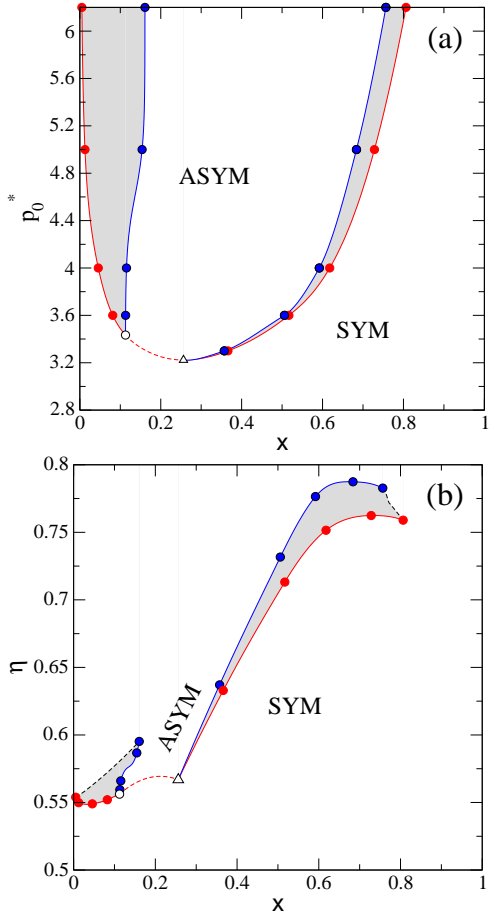

	\epsfig{file=Fig5a.eps,width=2.5in}
	\epsfig{file=Fig5b.eps,width=2.48in}
	\caption{Phase diagrams of a confined binary mixture of PHS with $\sigma_2/\sigma_1=1.5$ and 
$H/\sigma_1=2.6$. 
 (a) $p_0^*$ vs. ${\sf x}$. (b) $\eta$ vs. ${\sf x}$. Red and blue filled circles represent the coexisting 
	values of the SYM and ASYM states, respectively. Open circle represents the left-tricritical 
        point separating the coexisting binodals (solid lines) from the continuous phase-transition curve 
(dashed line). The open triangle represents the right-tricritical point. 
	}
	\label{fig5}
\end{figure}
The driving force for microsegregation is entropy. It is clear that, at close packing,  
two possible configurations of big clusters in the SYM phase are equally represented, 
i.e. big clusters containing big squares in contact with different walls are equally likely. 
In contrast, in the ASYM-phase this symmetry is broken, with one of the configurations overrepresented with 
respect to the other. Close packing can be attained by both ASYM and SYM phases, but the latter 
is more disordered in terms of mixing entropy and consequently has a lower free energy. However, far from 
close packing, when pressure is not too large (e.g. $p_0^*=4$), the situation can be different. 
Since big squares will be alternatively absorbed at both walls in the SYM phase, while the 
the space between big squares in contact with the same wall is moderately filled with small squares,
it is clear that it not possible for big squares to overpass each other: the motion
of small squares along the $x$-axis is severely restricted due to the jammed configuration of large particles. 
Thus the configurational entropy, related to the total number of allowed particle configurations, 
drops and consequently the free-energy increases as compared to that of the quasi-perfect ASYM-phase. 
In the latter big squares are not jammed (since most of them are adsorbed at the same wall) and 
therefore particles can move along the channel with much more freedom (the only constraint being
hard-core interactions with the lateral neighbors). 
Of course particles can also move along the $y$-axis, but they have similar 
freedom in both phases. 

\begin{figure}
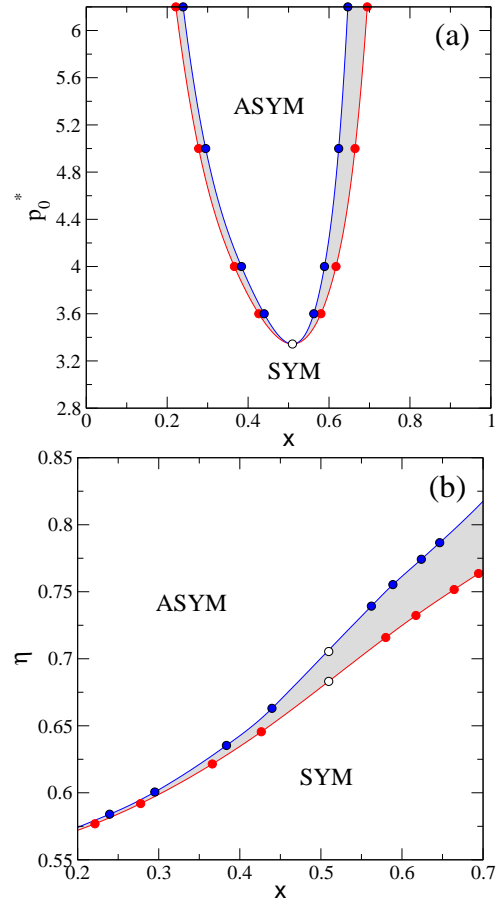

	\epsfig{file=Fig6a.eps,width=2.5in}
	\epsfig{file=Fig6b.eps,width=2.5in}
	\caption{Phase diagrams for the confined binary mixture of PHS with $\sigma_2/\sigma_1=1.5$ 
and $H/\sigma_1=2.8$. (a) $p_0^*$ vs. ${\sf x}$, and (b) $\eta$ vs. ${\sf x}$.
Red and blue circles represent the coexisting values corresponding to 
	SYM and ASYM states, respectively. The open circle indicates the azeotropic point.} 
	\label{fig6}
\end{figure}

\begin{figure*}
        \epsfig{file=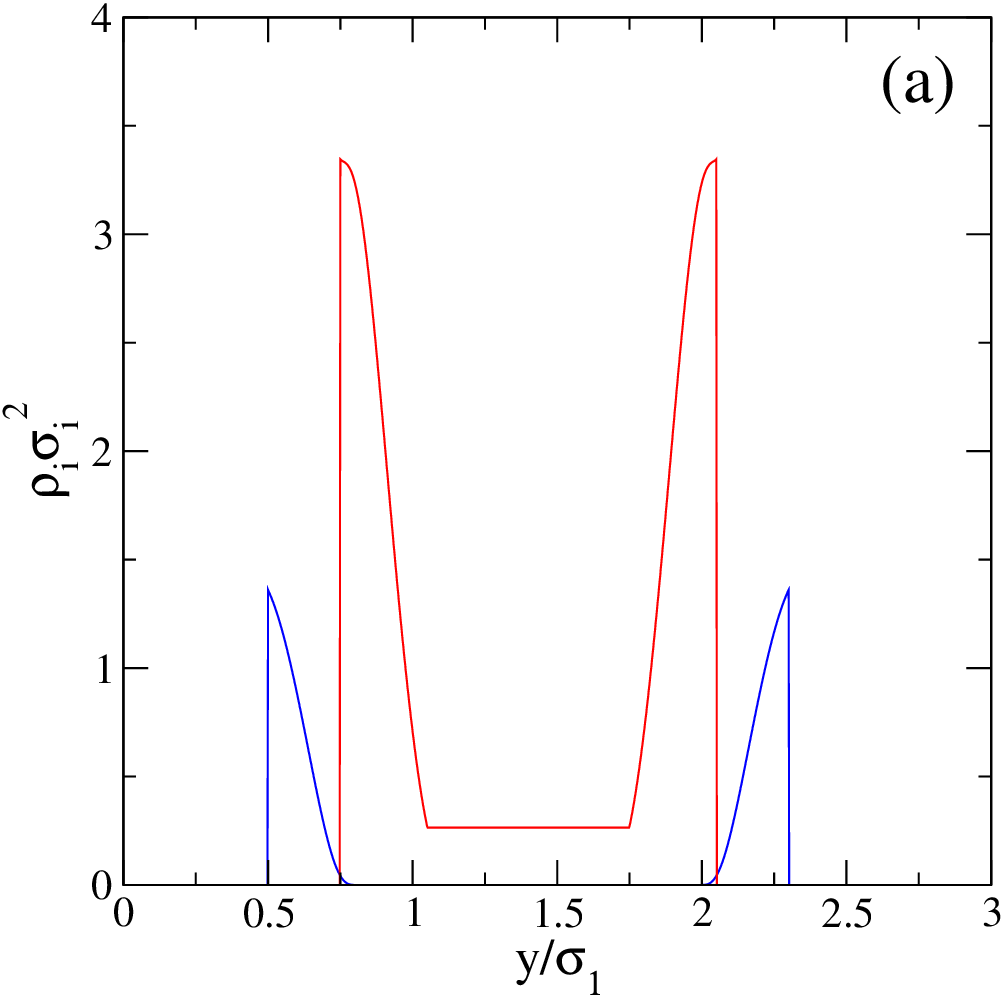,width=1.6in}
        \epsfig{file=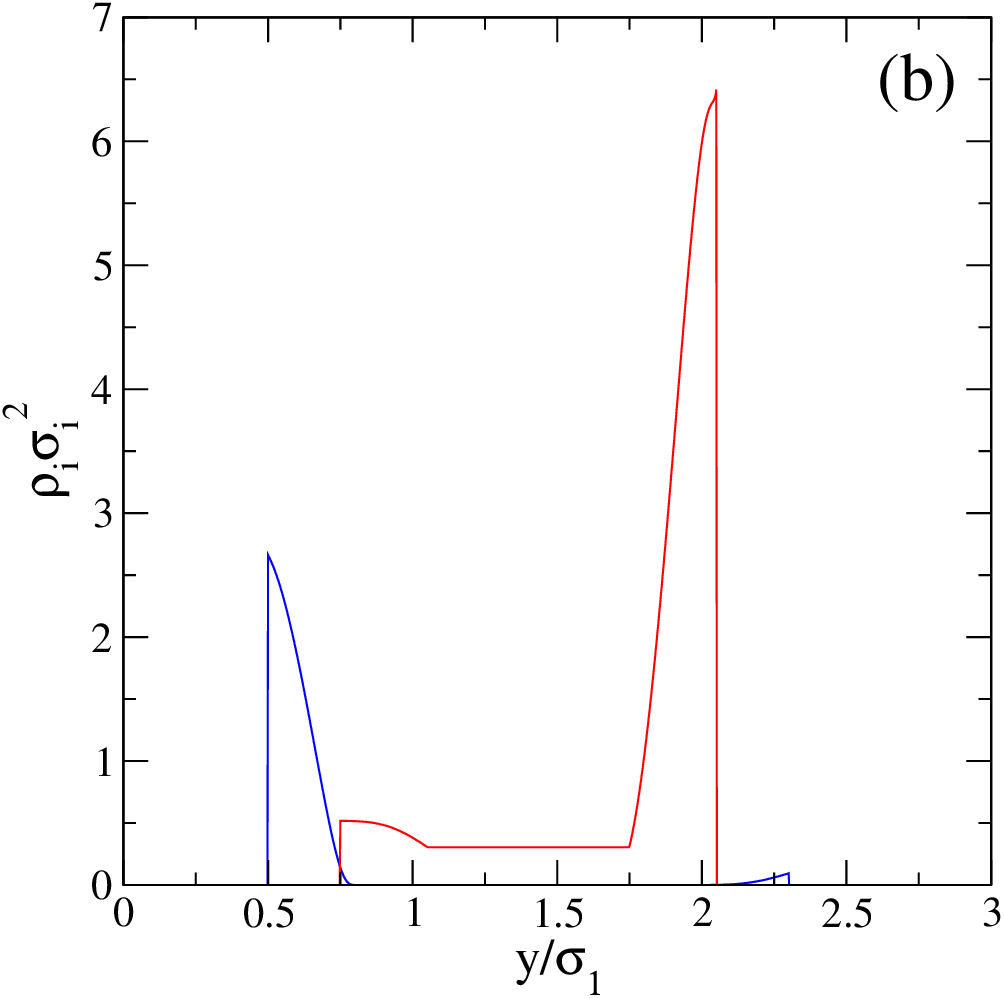,width=1.6in}
        \epsfig{file=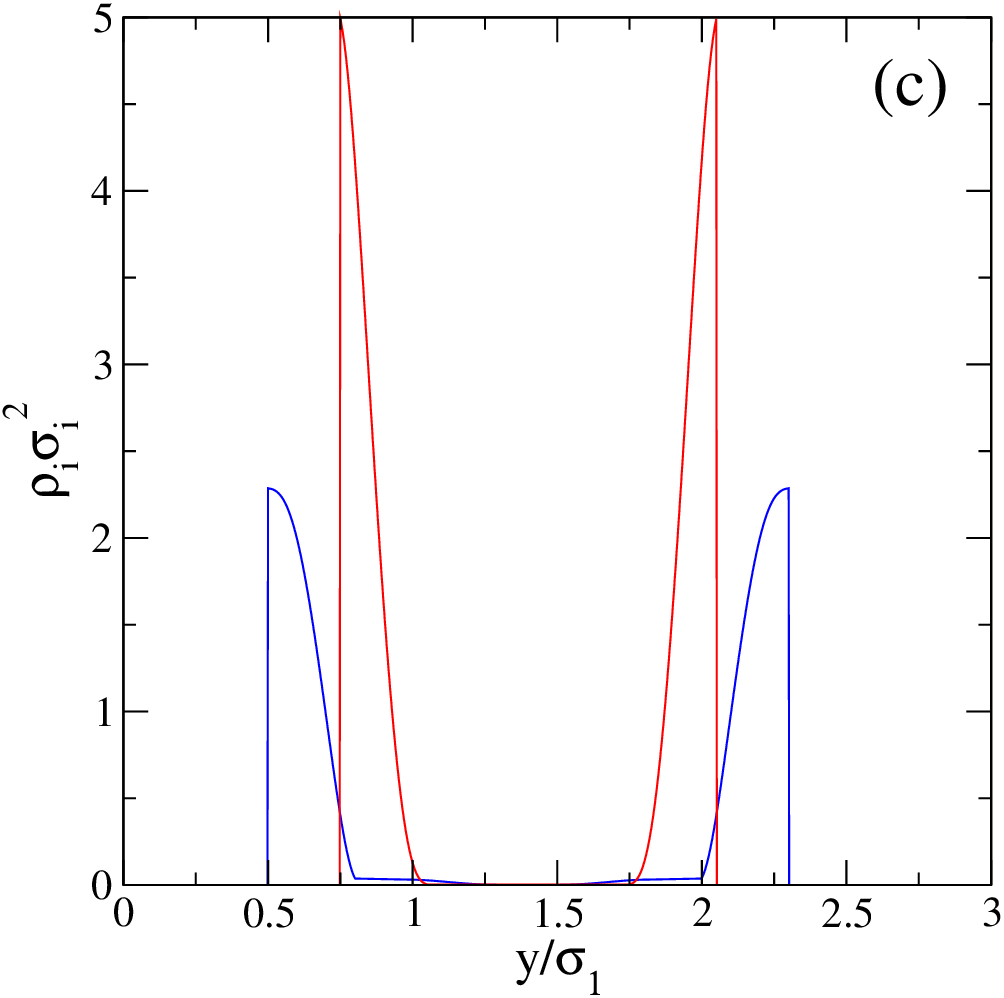,width=1.6in}
        \epsfig{file=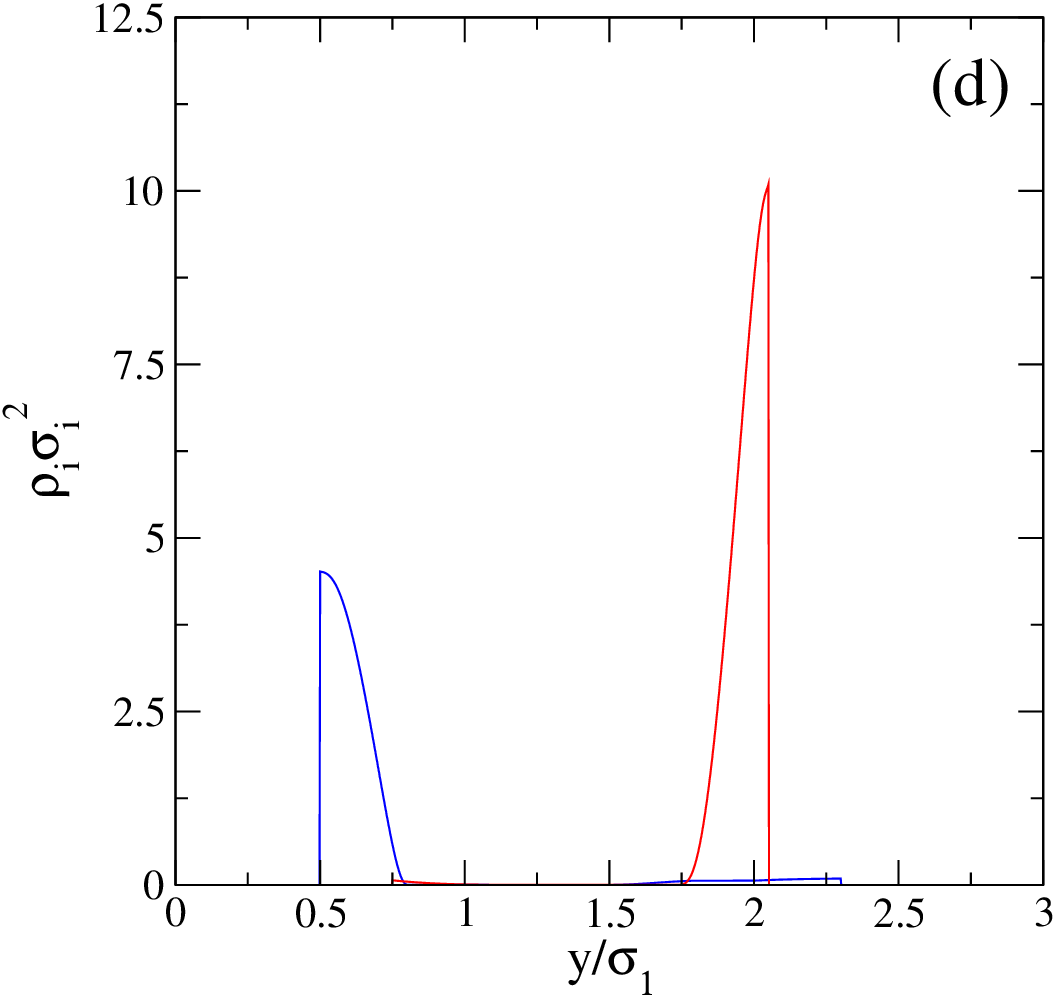,width=1.6in}

        \caption{Coexistence density profiles corresponding to the two first-order phase transitions shown
in Fig. \ref{fig6} for $\sigma_2/\sigma_1=1.5$, $H/\sigma_1=2.8$ and $p_0^*=4$. (a) and (b) correspond
to the low molar fraction region, whereas (c) and (d) refer to the high molar fraction region.
(a) and (c) are SYM phases, while (b) and (d) are ASYM phases. Blue and red curves
        correspond to scaled density profiles $\rho_1(y)\sigma_1^2$ and $\rho_2(y)\sigma_2^2$ of
        small and big species, respectively.}

	\label{fig7}
\end{figure*}

We performed coexistence calculations for several values of pressure to construct a phase diagram for 
$H/\sigma_1=2.6$. This is shown in Fig. \ref{fig5}, in the $p^*-{\sf x}$ and $\eta-{\sf x}$ planes. 
We see that the ASYM stability region is laterally bounded (in the ${\sf x}$ direction) 
by first-order SYM-ASYM and ASYM-SYM transitions lines. 
At low pressures the SYM-ASYM transition terminates in a 
left-tricritical point (open circle). From this point the 
transition becomes continuous. This line meets the  
binodals of the ASYM-SYM transition at the right-tricritical point 
(open triangle). Note the strong fractionation 
of the SYM-ASYM transition: the compositions of the coexisting phases 
are much more different than 
those of the ASYM-SYM transition. As more packed configurations are 
reached by increasing the amount of small squares, the phase diagram in 
the $\eta-{\sf x}$ plane [panel (b)] becomes highly asymmetric, i.e. 
there is a large difference in packing fraction values of the coexistence 
binodals at left and right of the end-critical point. 

The phase diagram for a wider pore width of $H/\sigma_1=2.8$, shown 
in Fig. \ref{fig6}, was also calculated.
In wider pores the entropically-driven microsegregation, resulting from 
particle-motion restrictions in jammed SYM-configurations, still operates,
but to a lesser extent because the transverse spatial
freedom of particles increase with $H/\sigma_1$. Thus the ratio 
between the gain in lateral free length (resulting from microsegregation)
and the transverse free length is lower. We should remind ourselves that 
configurational entropy competes with mixing entropy (which prevents 
microsegregated states). As a final result the region of ASYM-phase 
stability in the $p_0^*-{\sf x}$ phase diagram shrinks with 
$H/\sigma_1$, a fact that can be confirmed by looking at Fig. \ref{fig7} (a).  
It can be seen that for the highest pressure used ($p^*=6.2$) 
the stability interval in ${\sf x}$ of the ASYM-phase is now 
$\sim [0.23,0.65]$, smaller than $[0.15,0.75]$ (which corresponds 
to the $H/\sigma_1=2.6$-case). Another interesting feature of the 
phase diagram is the weaker character of the first-order SYM-ASYM transition
at the left of the azeotropic point (open circle). The azeotropic
character of the latter can be inferred from panel (b), which demonstrates
the existence of a coexisting gap in $\eta$ at this point, 
despite the fact that the composition of the coexisting phases is the same. 
The binodals are monotonically-increasing functions of ${\sf x}$, showing the 
higher packing inside the pore resulting from 
an increase in the number of small squares. 

Fig. \ref{fig7} depicts the 
four coexisting density profiles 
for this pore-width and with pressure fixed to $p_0^*=4$. 
The following results can be extracted: (i) density profiles are broadened 
compared to those for the thinner pore, and (ii) adsorption of big squares 
at the walls is increased: the heights of the central plateau in the density 
profile $\rho_2(y)$ [see (a) and (b)] are lower than those of 
Fig. \ref{fig3}(a) and (b). This effect can be understood in terms of 
the low values of coexisting compositions for the SYM-ASYM transition 
in the thin pore: There exists a large amount of big squares which do not 
contribute to the formation of the big clusters and they freely 
fluctuate between both walls.   

\begin{figure}
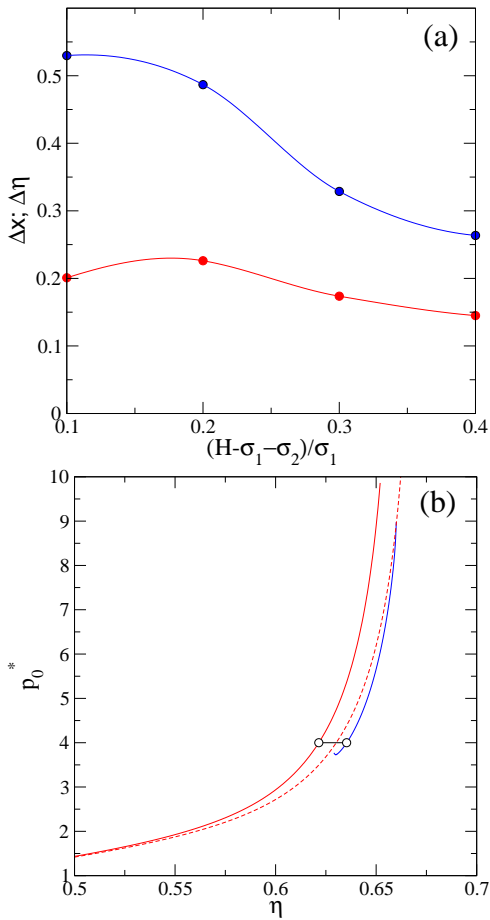

	\epsfig{file=Fig8a.eps,width=2.5in}
	\epsfig{file=Fig8b.eps,width=2.5in}
	\caption{(a) Differences in coexisting molar and packing
	fractions of ASYM phases, $\Delta {\sf x}\equiv {\sf x}^{(a,2)}-{\sf x}^{(a,1)}$ (blue) and
        $\Delta\eta\equiv \eta^{(a,2)}-\eta^{(a,1)}$ (red), as a function of the scaled free length
        $(H-\sigma_1-\sigma_2)/\sigma_1$, for a binary mixture with $\sigma_2/\sigma_1=1.5$
        and pressure $p_0^*=5$.
	(b) Equation of state (EOS) of a confined binary mixture with  
	$\sigma_2/\sigma_1=1.5$ and $H/\sigma_1=2.8$. The 
	red and blue solid lines correspond to SYM and ASYM states, with compositions fixed to
	their corresponding coexisting values at $p_0^*=4$ 
(i.e. ${\sf x}=0.36611$ and ${\sf x}=0.38352$ respectively). 
	Coexisting states are indicated by empty circles. 
	The EOS corresponding to a SYM phase with ${\sf x}=0.38352$ is shown by a dashed red curve.
Note that this curve intersects the blue line at high pressures, indicating that the ASYM states will
become unstable and an upper azeotropic point probably exists in the phase diagram. 
	}
	\label{fig8}
\end{figure}

The decrease in ASYM phase stability with pore-width $H/\sigma_1$ 
 at a fixed pressure (in particular for $p_0^*=4$)  
is confirmed in Fig. \ref{fig8} (a), where we plot the interval $\Delta {\sf x}={\sf x}^{a,2}-{\sf x}^{a,1}$ (with 
${\sf x}^{a,i}$ the coexisting values  of the left ($i=1$) and right ($i=2$) ASYM-binodals) 
in which stable ASYM solutions are found vs. the free transversal length 
$(H-\sigma_1-\sigma_2)/\sigma_1$. Also plotted are the difference in 
packing fractions
at these points ($\Delta \eta=\eta^{(a,2)}-\eta^{(a,1)}$) which does not 
change much but exhibits a maximun.

As mentioned before, the entropic mechanism that drives microsegregation 
at finite pressure does not operate at close packing because in this case 
(infinite pressure) the lateral free length that allows particle motion 
is absent, while mixing entropy favors the formation of SYM states. 
Therefore, at very high pressure, an upper azeotropic point is expected
in the phase diagram: we conjecture the existence of a finite region 
in the phase diagram where a reentrant ASYM-phase is stable. An indication 
that this could certainly be the case can be seen in 
Fig. \ref{fig8} (b), where we plot the EOS of the  
SYM (solid-red) and ASYM (solid-blue) phases for fixed values of compositions, 
${\sf x}=0.36611$ and ${\sf x}=0.38352$ respectively. These are the
coexisting values of the SYM-ASYM transition at $p_0^*=4$. 
The SYM and ASYM phases are stable in the intervals
$0<p_0^*<4$ and $4<p_0^*\lesssim 9$, respectively. Note how the EOS of 
a SYM phase with a fixed composition 
${\sf x}=0.38352$ (dashed red line) intersects the blue solid curve, which
indicates that for pressures $p^*\gtrsim 9$ the ASYM phase 
might loose stability with respect to the SYM phase. 
Unfortunately our numerical scheme to implement the DF minimization 
becomes unstable at these high pressures, and an alternative method, such
as a density-profile parameterization, is needed to validate this conjecture.

\begin{figure}
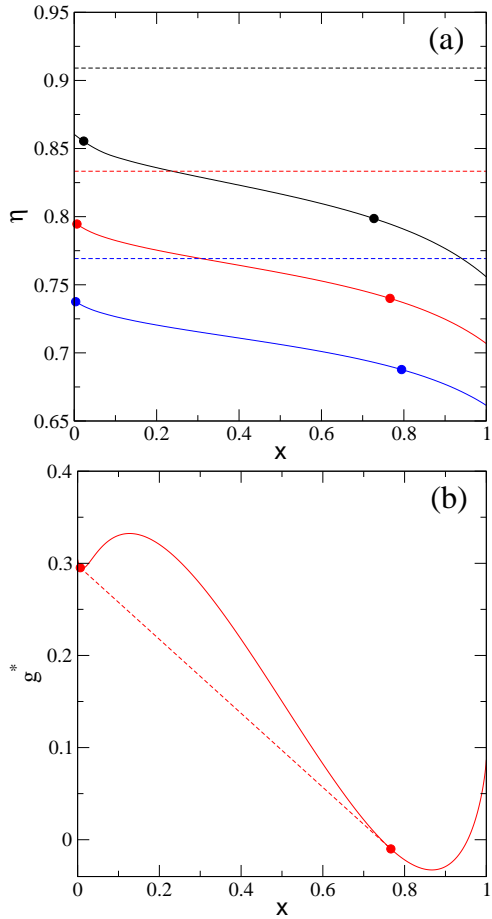

	\epsfig{file=Fig9a.eps,width=2.5in}
	\epsfig{file=Fig9b.eps,width=2.5in}
	\caption{(a) Packing fraction vs. molar fraction for a confined binary mixture of PHS with 
	$\sigma_2/\sigma_1=2$, $p_0^*=5$ and $H/\sigma_1=2.2$ (solid black) $2.4$ (solid red), and 
$2.6$ (solid blue).
	Close-packing values, $\eta_{\rm cp}$, for the same values of $H/\sigma_1$, are shown by dashed lines.
	(b) Scaled Gibbs free-energy per particle minus a straight line, 
$g^*\equiv \beta g-27.232+18.124 {\sf x}$, vs. 
	molar fraction for the same mixture and for $H/\sigma_1=2.4$. 
The solid circles joined with a dashed line indicate the coexistence values of ${\sf x}$.}
	\label{fig9}
\end{figure}

\subsection{The $\sigma_2/\sigma_1=2$ mixture}
\label{dos_uno}

\begin{figure}
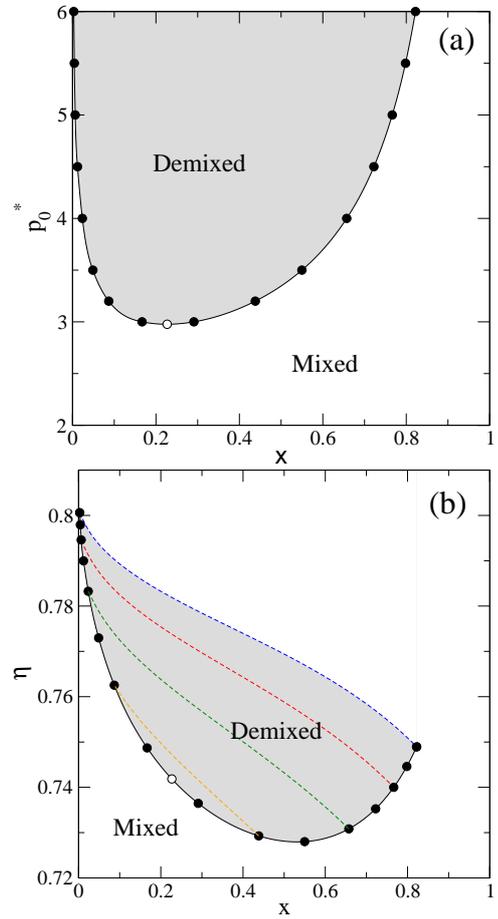

	\epsfig{file=Fig10a.eps,width=2.5in}
	\epsfig{file=Fig10b.eps,width=2.5in}
	\caption{Phase diagrams in the (a) pressure-composition, and (b) packing fraction-composition 
	planes of a confined binary mixture of PHS with $\sigma_2/\sigma_1=2$, and $H=2.4$. In (b)  
	four isobars, for $p_0^*=6$ (blue), 5 (red), 4 (green), and $3.2$ (orange) are shown 
        with dashed lines.}
	\label{fig10}
\end{figure}

\begin{figure}
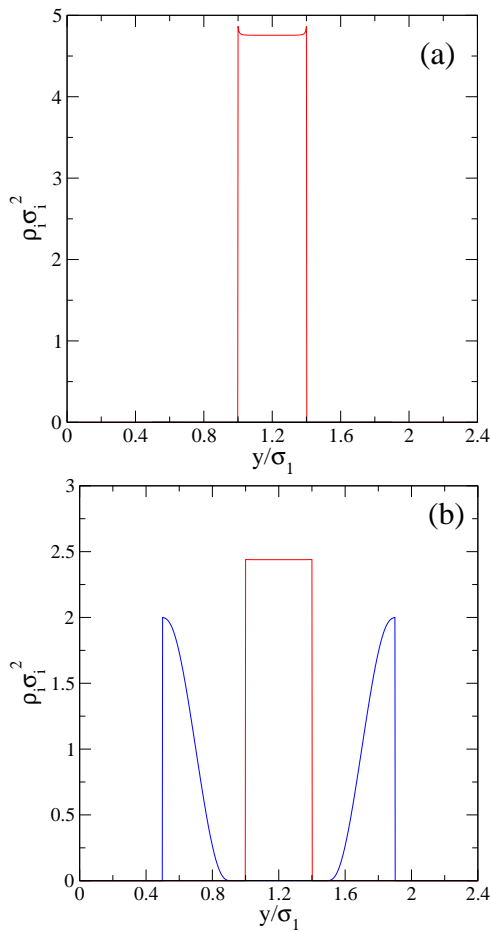

	\epsfig{file=Fig11a.eps,width=2.5in}
	\epsfig{file=Fig11b.eps,width=2.5in}
	\caption{Density profiles of a confined binary mixture with 
	$\sigma_2/\sigma_1=2$, $H/\sigma_1=2.4$ and $p_0^*=5$ corresponding to the coexisting phases 
	with (a) low and (b) high molar fractions.} 
	\label{fig11}
\end{figure}

\begin{figure}
	\epsfig{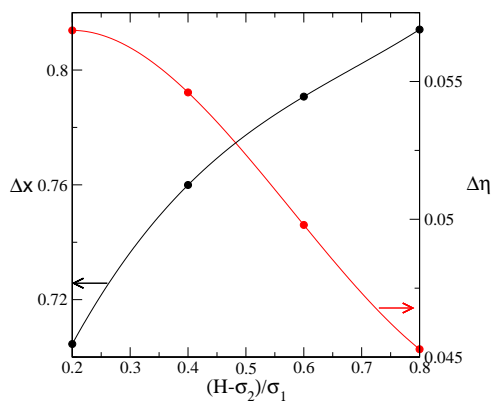}
	\caption{Differences in the coexisting molar fraction, 
$\Delta {\sf x}={\sf x}^{(2)}-{\sf x}^{(1)}$ (black), and packing
fraction, $\Delta \eta=\eta^{(2)}-\eta^{(1)}$ (red), of the demixed phases as 
        a function of the free length,
	$(H-\sigma_2)/\sigma_1$, for a binary mixture of confined PHS with $\sigma_2/\sigma_1=2$ and $p_0^*=5$.}
	\label{fig12}
\end{figure}

To find the close-packing configurations for $\sigma_2/\sigma_1=2$ 
we apply the same reasoning as before: The close-packed 
limit can be reached by joining ${\cal N}_{\rm b}$ 
big clusters (constituted by a single big square) with ${\cal N}_{\rm s}$ 
small clusters (formed by dimers of, perfectly 
aligned along $y$, small squares). The total area occupied  by both clusters 
is ${\cal N}_{\rm b}\sigma_2^2+2{\cal N}_{\rm s}\sigma_1^2$, whereas
the total occupied length along the channel is
${\cal N}_{\rm b}\sigma_2+{\cal N}_{\rm s}\sigma_1$. 
As the numbers $\{{\cal N}_{\rm b},{\cal N}_{\rm s}\}$  
fulfill the condition ${\cal N}_{\rm b}=(1-{\sf x}){\cal N}$ and ${\cal N}_{\rm s}={\sf x}{\cal N}/2$, we arrive at 
\begin{eqnarray}
	\eta_{\rm cp}&=&\frac{{\cal N}_{\rm b}\sigma_2^2+2{\cal N}_{\rm s}\sigma_1^2}{\left
	({\cal N}_{\rm b}\sigma_2+{\cal N}_{\rm s}
	\sigma_1\right)H}=\frac{(1-{\sf x})\left(\sigma_2/\sigma_1\right)^2+{\sf x}}
	{2(1-{\sf x})\sigma_2/\sigma_1+{\sf x}}\times \frac{2\sigma_1}{H}\nonumber\\&=&\frac{2\sigma_1}{H}.
\end{eqnarray}
The close-packing value does not depend on composition. 
In Fig. \ref{fig9} (a) these limits are shown for $H/\sigma_1=2.2$, 
2.4 and 2.6. Also the functions $\eta({\sf x})$ are plotted 
for the same values of 
pore widths as obtained from the DF minimization, by fixing the pressure 
to $p_0^*=5$. Clearly the packing fractions, monotonic decreasing functions 
of ${\sf x}$, do not change too much with composition as compared to the 
case $\sigma_2/\sigma_1=1.5$. The intervals of ${\sf x}$ between the solid 
circles represent the instability region in mixture composition with 
respect to demixing transitions. This behaviour can be confirmed by plotting
the Gibbs free-energy per particle (minus a straight line) 
$g^*$ vs. ${\sf x}$, as we do in panel (b) for $H/\sigma_1=2.4$ and $p_0^*=5$. 
Strong demixing between two confined phases, each one rich in one of species,
is confirmed. The phase separation has a clear lateral symmetry, 
i.e. both phases are separated along the channel with a
Gibbs-dividing interface perpendicular to the channel. This is a kind of 
macrosegregation, completely different to the microsegregation obtained 
before for the case $\sigma_2/\sigma_1=1.5$. However, the density profiles are 
now symmetric always.

\begin{figure}
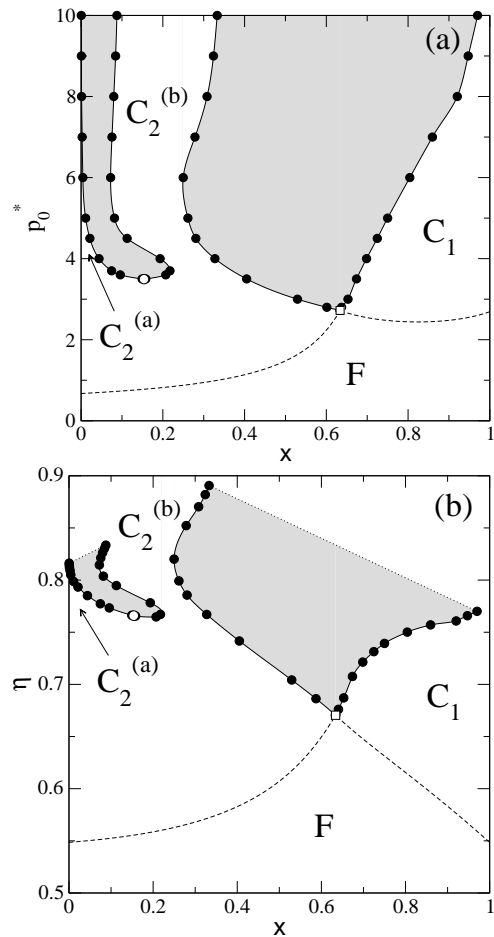

        \epsfig{file=Fig13a.eps,width=2.5in}
        \epsfig{file=Fig13b.eps,width=2.5in}
	\caption{(a) Phase diagram $p_0^*-{\sf x}$ 
	of a binary mixture of PHS with $\sigma_2/\sigma_1=2$, 
	$H/\sigma_1=2.4$ and PBC. 
Solid lines represent the coexisting binodals, whereas dashed lines 
indicate continuous phase transitions. 
	Regions of stability of fluid (F) and different columnar $\rm{C}^{(\alpha)}_{\it{i}}$ are 
	correspondingly shown. (b) The same phase diagram as in (a), but
in the $\eta-{\sf x}$ plane. Filled circles:
calculated binodal points. Open circle: critical point. Open square: tricritical point.}
	\label{fig13}
\end{figure}

\begin{figure*}
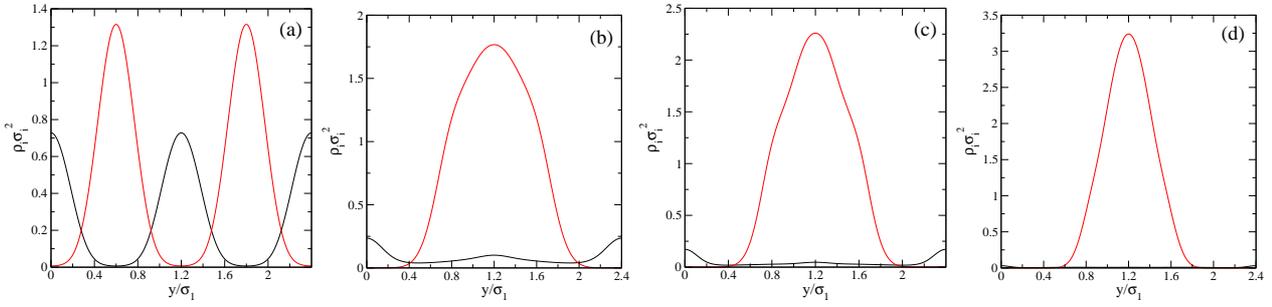

        \epsfig{file=Fig14a.eps,width=1.6in}
        \epsfig{file=Fig14b.eps,width=1.6in}
        \epsfig{file=Fig14c.eps,width=1.6in}
        \epsfig{file=Fig14d.eps,width=1.6in}
	\caption{Coexisting density profiles of a binary mixture with 
	$\sigma_2/\sigma_1=2$, $H/\sigma_1=2.4$, $p_0^*=4$ and PBC. 
	From (a) to (d) density profiles correspond to 
	${\sf x}^{(1)}>{\sf x}^{(2)}>{\sf x}^{(3)}>{\sf x}^{(4)}$, 
	i.e. the coexistence values of molar fractions for both demixing 
	transitions found in the phase diagram of Fig. \ref{fig13} at
the corresponding pressure.}
	\label{fig14}
\end{figure*}

To find the phase diagram, we have calculated the coexisting values of 
${\sf x}$ and $\eta$ for a set of different values of $p_0^*$, 
and for a fixed pore width $H/\sigma_1=2.4$, 
via the double-tangent construction of $g({\sf x})$. Phases diagrams
in $p^*-{\sf x}$ and $\eta-{\sf x}$ coordinates are plotted 
in Fig. \ref{fig10} (a) and (b) respectively.
Dashed lines in (b) correspond to different isobars inside the demixed region. 
The phase separation ends in a critical point (white circle), below which
the mixture is stable. As pressure is increased from that point, 
the coexisting phases become more similar to the confined one-component 
fluids. As an example of coexisting phases, Fig. \ref{fig11} shows 
the density profiles of small and big squares for the (a) low-${\sf x}$ and 
(b) large-${\sf x}$ coexisting phases, with $p_0^*=4$ and $H/\sigma_1=2.4$. 
Note that in panel (a) the density profile of small species is not visible at 
the scale of the figure, demonstrating the quasi-one-component character of the mixture.
We can see in panel (b) that, while big squares always fluctuate close to 
the center of the pore, small squares are strongly adsorbed at both walls.  

The phase separation once again is related to entropy. 
When both species are mixed,
e.g. when clusters formed by dimers of small squares are surrounded by big 
squares, lateral motion of small particles is strongly restricted because 
small and big species cannot overpass each other. Also, 
if one dimer of small particles is located between two big squares, 
motion of these highly constrained small squares entails the 
breaking of dimers, with a lowering in the local packing fraction. 
When the mixture is well separated, small squares can move in the lateral 
direction much more freely because the presence of other small particles
in front do not constrain their motion. Thus, the dimers can be continuously 
formed and destroyed without altering the local packing of particles.

\begin{figure}
	\epsfig{file=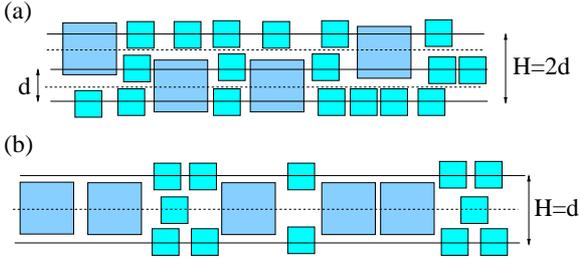,width=3.in}
	\caption{Schematic of particle configurations of two different columnar phases found in the phase 
	diagram with PBC: (a) ${\rm C}_1$, and (b) ${\rm C}_2^{(a,b)}$.}
	\label{fig15}
\end{figure}

An interesting issue is how the demixing transition depends 
on pore width $H$. The answer to this question is given by Fig. \ref{fig12},
where the demixing gaps in ${\sf x}$ (black) and $\eta$ (red) are plotted
as a function of the free length $(H-\sigma_2)/\sigma_1$ for a fixed pressure 
$p_0^*=5$.  As the pore becomes wider the demixing, in terms of 
fractionation, is stronger ($\Delta {\sf x}\equiv 
{\sf x}^{(2)}-{\sf x}^{(1)}$ is an increasing function of $H$),
while the gap in packing fraction decreases. 
The latter result is expected because packing of squares inside the pore 
is less effective as the pore becomes wider, so the two
coexistence values of $\eta$ decrease, to such an extent that the difference 
$\Delta\eta\equiv \eta^{(2)}-\eta^{(1)}$ is a monotonically
decreasing function of ${\sf x}$. However the relative gap, 
$\Delta\eta/\eta^{(1)}$, turns out to be constant with a value close to $0.07$. 
This interesting trend, namely a stronger demixing as $H$ increases, 
is opposite to that obtained for the $\sigma_2/\sigma_1=1.5$ mixture. 
As shown in the previous section, the microsegregation transition 
is enhanced when the pore becomes narrower.

To end this section, we comment on the relation between the 
phase behavior of the confined system and that of a similar system
subject to periodic boundary conditions (PBC). To investigate this, 
we have imposed the conditions $\rho_i(y+H)=\rho_i(y)$ on the density 
profiles, focusing on the binary mixture with $\sigma_2/\sigma_1=2$.  
The period $H$ was chosen to be equal to the pore width of one of the
mixtures analysed previously, i.e. $H/\sigma_1=2.4$.
We did not minimize the DF with respect to $H$, with the aim to making
the comparison of the two results meaningful. Consequently the phase diagram 
presented below is not the bulk one. Note that PBC are normally used to 
mimic bulk phase behavior when the system is infinite along the $y$ 
direction, with the inhomogeneous phase being periodic along the same 
direction. But the same condition can also describe a finite system of 
dimension $H$ in the $y$ direction. Unfortunately the DF is unable to
distinguish both situations, which is a strong drawback of this 
theoretical tool. Obviously, if a DF based on the two-body probability density, 
instead of the one-body density, could be constructed, it would certainly 
contain information on the finiteness of the system along $y$. Therefore,
at present, results from the (one-body density-based) DF and the TMM 
applied to the study of systems with PBC cannot be compared \cite{Gurin7}. 

Fig. \ref{fig13} shows the phase diagram as obtained from DF minimization. 
The dashed lines represent continuous transitions 
between a fluid of PHS and 
a periodic columnar phase with period $H/\sigma_1=2.4$. 
The latter was 
calculated by searching for the divergence of the 
structure-factor inverse matrix, as described in Sec. \ref{app_sf}. The solid lines 
(which join the calculated points) are the coexisting binodals of the 
demixing transitions.
For relatively high composition, ${\sf x}\gtrsim 0.7$, and fixed pressure $p_0^*=4$, we find that the stable phase is the so-called 
${\rm C}_1$ columnar phase, formed by two layers of small squares (of period $d\equiv H/2=1.2\sigma_1$) where 
the centers of mass of big squares occupy interstitial 
positions between the layers and the density 
profiles are out of phase: $\rho_2(y)=\rho_1(y+d/2)$. 
See Fig. \ref{fig14} (a), where these density profiles are
plotted, and Fig. \ref{fig15} (a) for a sketch of particle configurations. Note that big squares 
intersect the two adjacent layers formed by small squares. 
As ${\sf x}$ is decreased this phase looses stability at 
${\sf x}\sim 0.7$, and the 
system exhibits strong demixing to the so-called ${\rm C}_2^{(b)}$-columnar phase,
with a composition ${\sf x}\sim 0.3$ and formed by layers of big squares with small squares mostly microsegregated 
at the interstitials [see Fig. \ref{fig14} (b) for the density profiles and Fig. \ref{fig15} (b) for a
sketch of particle configurations]. Now the periodicity is $d=H$. By further decreasing ${\sf x}$ it is found that this
phase is stable up to ${\sf x}\sim 0.2$, where a new phase transition takes place to the so-called ${\rm C}_2^{(a)}$ 
phase. This is very similar in structure to the ${\rm C}_2^{(b)}$ phase, but the former exhibits a domed-like 
density profile for the big squares 
[see Fig. \ref{fig14} (c)], while the latter has the usual 
sharply-peaked form 
[see Fig. \ref{fig14} (d)], with a small amount of small 
squares located at the interstitials. The 
${\rm C}_2^{(a)}-{\rm C}_2^{(b)}$ and ${\rm C}_1-{\rm C}_2^{(b)}$ transitions end in critical and tricritical points, 
respectively. The main conclusion drawn from these results is that scenarios, 
strong demixing and microsegregation, 
are also present in a PHS fluid subject to PBC. This is an indication that the
the bulk phase diagram will also contain these two features.

Crystalline phases (where both density profiles depend on both spatial coordinates) 
were not included in our study. At high pressure crystals will certainly become 
more stable than the exotic one-dimensional profiles we have found in the region of 
stability of $C_2^{(b)}$ at very high pressures (not shown here).

\section{Discussion and Conclusions}
\label{conclusions}

We have used the DF formalism, based on the FMT, to study the packing properties of extremely 
confined mixtures of PHS in a slit pore. Two types of mixtures have been analysed in detail 
by appropriately choosing particle sizes and pore width. In a first study, parameters were tuned
to avoid configurations where two big squares are located opposite to each other while dimers of 
one big and one small or two small squares, but not three of them, can fit into the channel. 
In a second study parameters were arranges so as to avoid dimers formed by one big and one small 
square to fit into the channel, while two small squares can fit. 
We have shown that the theory predicts micro- and macrosegregation phase transitions for the first 
and second mixture, respectively.  
Using the Gibbs free energy potential for a set of fixed pressures,
coexisting packing and molar fractions were calculated via a double-tangent construction. 
Thus the first-order character of phase transitions at most pressures was identified, 
and boundaries of stability regions for mixed and micro (macro)-demixed states were traced out. 
All phase transitions found have an entropic character, which is ultimately related to the
jammed configurations of particles. These configurations arise when two big 
squares are close to each other and, at the same time and for the first mixture, they are 
symmetrically adsorbed at both walls.
The jammed configurations severely restrict lateral motion of small particles (thus decreasing the 
configurational entropy), and these can explore a limited space as compared to that in macro- or 
microsegregated mixtures. Finally, by imposing PBC, we showed that
demixing transitions between different columnar phases 
also take place in systems without external potentials restricting particle positions.   
In this case all demixed phases found also have a microsegregated structure, with one of the 
species forming the main columns and the others occupying the interstitial regions.

A comment on the real nature of phase transitions obtained here for the confined system  
is in order. Exact calculations using the TMM for one-component hard disks, squares, rectangles 
or rhombuses confined in a slit geometry, with at most two layers of particles, show that these  
$1+\epsilon$-dimensional systems do not exhibit true phase transitions \cite{Gurin5,Gurin4,Gurin2,Gurin1}. However their structural properties can dramatically change as pressure is increased. 
This behaviour is usually associated with a peculiar shape of the EOS which, 
under certain conditions, contains a plateau-like segment and a corresponding sharp
peak is visible in the heat capacity. With these results in mind, 
our analysis based on the (mean-field) DFT suggests that changes in particle configurations, 
driven by entropic forces, are adequately described by the theory, while the corresponding
phase transitions are not. Our claim is that, for high enough pressures, the two confined mixtures
studied here will contain an important number of large micro- and macrosegregated clusters, respectively. 
Although these clusters can symmetrically adopt two configurations in the case of the first mixture 
studied, their presence can be confirmed by calculating the two-body particle correlation function
using TMM. 

The most important result from our mean-field model, as applied to the second confined mixture 
(with $\sigma_2/\sigma_1=2$), consists in the prediction of a lateral demixing transition 
between two phases, each one rich in one species, at high enough pressures. 
We should bear in mind that, once monomers or dimers of small particles become located between 
two big squares, they will not be able to escape from the cage formed by big particles,
due to the impossibility that small and big squares can pass each other. Thus, if an equimolar 
mixture is initially prepared in a configuration where particles are randomly positioned 
(and consequently there is a high probability to find many small particles between the large ones) 
at high enough packing fraction, the mixture will become thermodynamically unstable with respect 
to phase segregation. However, the system will be unable to reach equilibrium (with two phases 
laterally segregated), as predicted from the thermodynamical analysis, due to severe particle jamming. 
These equilibrium states are not accessible in our strictly two-dimensional system. 
However our system could still approximately describe an experimental realization consisting of a 
colloidal binary mixture of hard cubes sedimented in a container with 
a nano-sculpted bottom surface under micro-gravity conditions. The surface could be nano-structured 
with quasi-2D channels, such that monolayers of sedimented cubes inside each channel were in contact 
with a ``bath'' of particles. The fact that particles can now enter or escape from the channel 
avoids jamming effect and the mixture could reach a final state with two segregated ``phases'' 
inside the channels. Note that the system would not be strictly two-dimensional as the channels 
would interact with the bulk regions in a nontrivial manner.

\appendix

\section{Uniform phases at bulk}
\label{uniform}

For uniform densities $\rho_i$, the ideal and excess parts of the free-energy density, and the fluid
pressure, are given by
\begin{eqnarray}
	&&\Phi_{\rm id}=\rho\left(\log \rho -1 +\sum_i {\sf x}_i\log {\sf x}_i\right),\nonumber\\
        &&\Phi_{\rm exc}=-\log(1-\eta)+\frac{n_1^2}{1-\eta},\nonumber\\
        &&\beta p=\frac{\rho}{1-\eta}+\frac{n_1^2}{(1-\eta)^2},
\end{eqnarray}
where we have defined $n_1=n_{1x}=n_{1y}=\sum_i \rho_i \sigma_i$.
Thus, the Gibbs free-energy per particle for a fixed pressure $p_0$
in reduced thermal units can be calculated as
\begin{eqnarray}
        && g\equiv \frac{\beta G}{N}=\frac{\Phi_{\rm id}+\Phi_{\rm exc}+\beta p_0}{\rho}\nonumber\\
	&&=\log\rho^*-1+\sum_i {\sf x}_i\log {\sf x}_i-\log(1-\rho^*s_2)\nonumber\\&&+\frac{\rho^*s_1^2}{1-\rho^*s_2}
        +\frac{p_0^*}{\rho^*},\label{gibbs2}
\end{eqnarray}
where we have defined the dimensionless number density and pressures as $\rho^*=\rho\sigma_1^2$
and $p_0^*=\beta p_0\sigma_1^2$. Also we have defined the quantities
\begin{eqnarray}
	s_m\equiv\sum_i {\sf x}_i\left(\frac{\sigma_i}{\sigma_1}\right)^m
	=\lambda^m-\left(\lambda^m-1\right){\sf x},
\end{eqnarray}
for $m=\{1,2\}$, with the aspect ratio of the mixture defined as $\lambda=\sigma_2/\sigma_1$.
From the constant pressure condition we can calculate
$\rho^*$ as a function of the composition which results in
\begin{eqnarray}
        \rho^*=\frac{1+2p_0^*s_2-\sqrt{1+4p_0^*s_1^2}}{2\left(p_0^*s_2^2+s_2-s_1^2\right)}.
        \label{larho2}
\end{eqnarray}
After substitution of (\ref{larho2}) into Eq. (\ref{gibbs2}) we obtain
\begin{eqnarray}
	&&g({\sf x})
=p_0^*s_2+\sum_i {\sf x}_i\log {\sf x}_i+\log\left(\frac{\sqrt{1+4p_0^*s_1^2}-1}{2s_1^2}\right)
	\nonumber\\&& +\sqrt{1+4p_0^*s_1^2}-1.
\end{eqnarray}
It can be easily shown that the second derivative of $g({\sf x})$ with respect to ${\sf x}$ gives the condition
\begin{eqnarray}
	\frac{d^2g}{d{\sf x}^2}({\sf x})=\frac{1}{{\sf x}(1-{\sf x})}+\left(\frac{\lambda-1}{s_1}\right)^2
	\left[1-\frac{1}{\sqrt{1+4p_0^*s_1^2}}\right]>0,\nonumber\\
\end{eqnarray}
$\forall \ \{{\sf x},p_0^*,\lambda\}$.
Thus the Gibbs free-energy per particle is always a convex function of composition
and consequently we can draw the important conclusion that the PHS fluid is always
stable with respect to phase separation.

\section{Spinodal instabilities to bulk non-uniform phases}
\label{app_sf}

We consider here the instability of the fluid phase with respect to inhomogeneities in one direction, say $y$.
We need to calculate the direct correlation functions
\begin{eqnarray}
	&&-c_{ij}(y-y')=\frac{\delta^2\beta {\cal F}[\{\rho_i\}]}{\delta \rho_i(y) 
	\delta \rho_j(y')}\nonumber\\&&=\sum_{\alpha\beta}\Phi_{\alpha\beta} 
        \left[\omega^{(\alpha)}_i\ast\omega^{(\beta)}_j\right](y-y'), 
\end{eqnarray}
where we have defined
$\displaystyle{\Phi_{\alpha\beta}=\frac{\partial^2 \Phi_{\rm exc}}{\partial n_{\alpha}\partial n_{\beta}}}$
in the uniform limit, and
the symbol $\ast$ stands for convolution. The weighting functions $\omega^{(\alpha)}_i(y)$
are those which define the weighted densities through convolutions:
\begin{eqnarray}
        n_{\alpha}(y)=\sum_i \left[\rho_i\ast\omega^{(\alpha)}_i\right](y).
\end{eqnarray}
They have the explicit forms
\begin{eqnarray}
	&&\omega^{(0)}_i(y)=\frac{1}{2}\delta\left(\frac{\sigma_i}{2}-|y|\right),\nonumber \\ 
	&&\omega^{(2)}_i(y)=\sigma_i\Theta\left(\frac{\sigma_i}{2}-|y|\right),\nonumber\\
	&&\omega^{(1x)}_i(y)=\frac{\sigma_i}{2}\delta\left(\frac{\sigma_i}{2}-|y|\right),\nonumber\\
	&&\omega^{(1y)}_i(y)=\Theta\left(\frac{\sigma_i}{2}-|y|\right),
\end{eqnarray}
with $\delta(y)$ and $\Theta(y)$ the Dirac-delta and Heaviside functions, respectively.
The Fourier transforms of the functions $c_{ij}(y)$ give
\begin{eqnarray}
        -\hat{c}_{ij}(q)=\sum_{\alpha,\beta} \Phi_{\alpha\beta} 
        \hat{\omega}_i^{(\alpha)}(q) \hat{\omega}_j^{(\beta)}(q),
\end{eqnarray}
where $q$ is the wave number, while
\begin{eqnarray}
	&&\hat{\omega}_i^{(0)}(q)=\cos\left(\frac{q\sigma_i}{2}\right),\nonumber\\ 
	&&\hat{\omega}_i^{(2)}(q)=\frac{2\sigma_i}{q}\sin\left(\frac{q\sigma_i}{2}\right),\nonumber\\
	&&\hat{\omega}_i^{(1x)}(q)=\sigma_i\cos\left(\frac{q\sigma_i}{2}\right),\nonumber\\
	&&\hat{\omega}_i^{(1y)}(q)=\frac{2}{q}\sin\left(\frac{q\sigma_i}{2}\right).
\end{eqnarray}
After a little algebra we arrive at
\begin{eqnarray}
        &&-\hat{c}_{ij}(q)=\frac{2}{1-\eta}\left(\sigma_i+\sigma_j +\frac{n_1}{1-\eta}\sigma_i\sigma_j\right)\nonumber\\
	&&\times \frac{\sin[q(\sigma_i+\sigma_j)/2]}{q}
        \nonumber\\
        &&+\frac{4}{(1-\eta)^2}\left[n_1(\sigma_i+\sigma_j)+\left(\rho+\frac{2n_1^2}{1-\eta}\right)\sigma_i\sigma_j\right]
        \nonumber\\
	&&\times \frac{\sin(q\sigma_i/2)\sin(q\sigma_j/2)}{q^2}.
\end{eqnarray}
The determinant of the inverse structure factor matrix,
\begin{eqnarray}
        S^{-1}_{ij}(q,\eta)=\delta_{ij}-\sqrt{\rho_i\rho_j}\hat{c}_{ij}(q),
\end{eqnarray}
can be calculated as
\begin{eqnarray}
	&&{\cal S}(q,\eta)\equiv \text{det} \left[ S^{-1}_{ij}\right](q)= 
        1-\rho_1\hat{c}_{11}(q)-\rho_2\hat{c}_{22}(q)\nonumber\\
	&&+\rho_1\rho_2
        \left[\hat{c}_{11}(q)\hat{c}_{22}(q)-\hat{c}_{12}(q)^2\right]
        \label{sondos}
\end{eqnarray}
Thus, the minimum value of $\eta$ for which the equations
\begin{eqnarray}
        {\cal S}(q,\eta)=0, \quad \frac{\partial {\cal S}}{\partial q}(q,\eta)=0,
	\label{hay_dos}
\end{eqnarray}
are fulfilled at the absolute minimum of ${\cal S}(q,\eta)$ as a function of $q$ provides the
values $q^*$ and $\eta^*$ at bifurcation. These calculations are done by fixing the molar fraction
of the mixture ${\sf x}$. Varying ${\sf x}$ and solving Eqs. (\ref{sondos})  we find the spinodal curve 
$\eta^*({\sf x})$ and
the periodicity of the non-uniform phases $d({\sf x})\equiv 2\pi/q^*({\sf x})$.
Note that if we fix the periodicity of the density profiles as $\rho_i(y+d)=\rho_i(y)$, as we have done at the end of 
Sec. \ref{dos_uno}, we need to solve only the first Eq. in (\ref{hay_dos}).

\acknowledgements

Financial support under grants FIS2015-66523-P, PGC2018-096606-B-100 and FIS2017-86007-C3-1-P
from Ministerio de Econom\'{\i}a, Industria y Competitividad (MINECO) of Spain is acknowledged. We gratefully acknowledge fruitful discussions with
P. Gurin and S. Varga.

\end{document}